\documentclass[submission,copyright,creativecommons]{eptcs}
\providecommand{\event}{PDMC 2011}
\usepackage{breakurl}
\usepackage[utf8]{inputenc}

\usepackage{nccbbb}
\usepackage{amssymb,
	    amsmath,
	    mathrsfs,
	    url,
	    color,
	    ifpdf,
	    setspace,
	    semantic,
	    xspace,
	    pstricks,
	    pst-node,
	    booktabs,
	    wrapfig,
	    colortbl}
\usepackage[final]{graphicx}
\usepackage{subfig}
\usepackage{latexsym}
\usepackage{hyperref}
\usepackage{psfrag}
\usepackage{wrapfig}
\usepackage{ifthen}
\usepackage{floatfig}
\usepackage[english]{babel}
\usepackage{listings}
\definecolor{darkgreen}{rgb}{0,0.5,0}
\definecolor{darkred}{rgb}{0.5,0,0}
\lstdefinelanguage{Uppaal}{ 
  basicstyle=\small\sffamily, 
  keywords={after update,assign,before update,break,case,const,continue,
     default,else,enum,for,guard,if,meta,process,progress,return,select,
     state,sync,switch,trans,system,while},
  keywords={[2]broadcast,bool,clock,chan,commit,init,int,scalar,struct,
    typedef,urgent,void}, keywordstyle={[2]\bfseries},
  keywords={[3]false,true}, keywordstyle={[3]\bfseries},
  comment=[l]{//}, morecomment=[s]{/*}{*/}, 
  morecomment=[s][\itshape\bfseries]{/**}{*/},
  commentstyle=\itshape, 
  tabsize=4, 
  captionpos=b, 
  frameround=tttf, frame=trbl, xleftmargin=7mm,xrightmargin=2mm,
  escapechar=@ 
}
\lstdefinelanguage[GUI]{Uppaal}[]{Uppaal}{ 
  keywordstyle={[2]\color{darkgreen}}, 
  keywordstyle={[3]\color{magenta}},
  commentstyle={\color{darkred}\itshape}, 
  morecomment=[s][\color{darkred}\itshape\bfseries]{/**}{*/},
  breaklines=true, breakatwhitespace=true 
}

\usepackage{amssymb}
\usepackage{xspace}
\usepackage{textcomp} 
\usepackage{calc}

\DeclareMathAlphabet\mathscra{T1}{hlcw}{m}{it}

\let\then\iftrue
{\catcode`\!=8 
 \catcode`\Q=3 
 \catcode`\@=11
 \long\gdef\ifgiven#1\then{\Ifbl@nk#1QQQ\empty!}
 \long\gdef\ifblank#1\then{\Ifbl@nk#1QQ..!}
 \long\gdef\Ifbl@nk#1#2Q#3!{\ifx#3}
 \long\gdef\ifnull#1\then{\IfN@LL#1* {#1}!}
 \long\gdef\IfN@LL#1 #2!{\ifblank{#2}\then}}

    {\bigskip\noindent\begin{tabular}{|p{\textwidth}}\it\small\strut}%
    {\strut\end{tabular}\par\bigskip}

    {\begingroup\vskip-2ex\begin{small}\noindent}%
    {\end{small}\vskip1.5mm\endgroup}

    {\begingroup\small\begin{equation}}%
    {\end{equation}\endgroup}
\newcommand*{\stignore}[1]{}

\newcommand{\wordsp}[1]{\textit{#1}}

\newcommand*{\mktype}[2]{%
  \expandafter\newcommand\csname #1sym\endcsname%
      {{\textsf{#2}}}%
  \expandafter\newcommand\csname #1name\endcsname%
      {{\csname #1sym\endcsname}\xspace}%
  \expandafter\newcommand\csname #1\endcsname%
      {{\csname #1sym\endcsname}\xspace}%
    }

\newcommand*{\mkkeyword}[2]{%
  \expandafter\newcommand\csname #1sym\endcsname%
      {{\ensuremath{#2}\expandafter\index{#2@\protect{\csname #1sym\endcsname}}}}%
  \expandafter\newcommand\csname #1name\endcsname%
      {{\csname #1sym\endcsname}\xspace}%
  \expandafter\newcommand\csname #1\endcsname%
      {{\csname #1sym\endcsname}\xspace}%
    }

\newcommand*{\mkfunc}[3]{%
  \expandafter\newcommand\csname #1sym\endcsname%
  {\ensuremath{#2}\index{#1@\protect{\csname #1sym\endcsname}}}%
  \expandafter\newcommand\csname #1name\endcsname%
  {\csname #1sym\endcsname\xspace}%
  \expandafter\newcommand\csname #1\endcsname[1]%
  {\ensuremath{\csname #1sym\endcsname #3{##1}}\xspace}%
}

\newcommand{\encloseinpar}[1]{(#1)}

\newcommand{\encloseinsembra}[1]{|[#1|]}
\newcommand{\encloseinangles}[1]{\langle #1\rangle}

\newcommand{\mksubscript}[1]{_{#1}}
\newcommand{\donotenclose}[1]{#1}
\newcommand{\followwithbang}[1]{#1{\ensuremath{\scriptstyle !}}}
\newcommand{\followwithqstn}[1]{#1{\ensuremath{\scriptstyle ?}}}
\newcommand{\followwithnot}[1]{#1{\raisebox{-0.8ex}[0mm][.1ex]{\ensuremath{\scriptstyle \!\not\,\,}}}}
\newcommand{\followwithast}[1]{#1^\ast}

\newcommand*{\mkfuncn}[2]{\mkfunc{#1}{#2}{\donotenclose}}    
\newcommand*{\mkfuncp}[2]{\mkfunc{#1}{#2}{\encloseinpar}}    
\newcommand*{\mkfuncs}[2]{\mkfunc{#1}{#2}{\encloseinsembra}} 
\newcommand*{\mkfuncsub}[2]{\mkfunc{#1}{#2}{\mksubscript}}   

\newcommand*{\mkbinaryrel}[2]{%
  \expandafter\newcommand\csname#1sym\endcsname%
  {{\ensuremath{#2}}}%
  \expandafter\newcommand\csname not#1sym\endcsname%
  {{\ensuremath{\not{#2}}}}%
  \expandafter\newcommand\csname #1name\endcsname%
  {\csname #1sym\endcsname\xspace}%
  \expandafter\newcommand\csname not#1name\endcsname%
  {\csname not#1sym\endcsname\xspace}%
  \expandafter\newcommand\csname #1\endcsname[2]%
  {{\ensuremath{##1#2##2}}}%
  \expandafter\newcommand\csname not#1\endcsname[2]%
  {{\ensuremath{##1 \not#2  ##2}}}%
}

\newcommand*{\mkternaryrel}[2]{%
  \expandafter\newcommand\csname #1sym\endcsname%
  {\ensuremath{#2}}%
  \expandafter\newcommand\csname not#1sym\endcsname%
  {\ensuremath{\not #2}}%
  \expandafter\newcommand\csname #1name\endcsname%
  {\csname #1sym\endcsname\xspace}%
  \expandafter\newcommand\csname not#1name\endcsname%
  {\csname not#1sym\endcsname\xspace}%
  \expandafter\newcommand\csname #1\endcsname[3]%
  {\ensuremath{##1 \csname #1sym\endcsname_{##2} ##3}}%
  \expandafter\newcommand\csname not#1\endcsname[3]%
  {\ensuremath{##1 \not\csname #1sym\endcsname_{##2} ##3}}%
}

\newcommand*{\mkternaryrelraised}[2]{%
  \expandafter\newcommand\csname #1sym\endcsname%
  {\ensuremath{#2}}%
  \expandafter\newcommand\csname not#1sym\endcsname%
  {\ensuremath{\not #2}}%
  \expandafter\newcommand\csname #1name\endcsname%
  {\csname #1sym\endcsname\xspace}%
  \expandafter\newcommand\csname not#1name\endcsname%
  {\csname not#1sym\endcsname\xspace}%
  \expandafter\newcommand\csname #1\endcsname[3]%
  {\ensuremath{##1 \csname #1sym\endcsname{}^{##2} ##3}}%
  \expandafter\newcommand\csname not#1\endcsname[3]%
  {\ensuremath{##1 \not\csname #1sym\endcsname{}^{##2} ##3}}%
}

\newcommand*{\mkcomrel}[5]{%
  \ifgiven{#4}\then%
    \expandafter\newcommand\csname #1sym\endcsname%
    {\raisebox{.2ex}{\ensuremath{\xrightarrow[{\raisebox{.5ex}[0mm][.1ex]{\tiny #4}}]%
          {#3{}}}}{}%
    \index{#4@\protect{\csname #1sym\endcsname}}}%
    \expandafter\newcommand\csname #1name\endcsname%
    {\csname #1sym\endcsname\xspace}%
    \expandafter\newcommand\csname #1\endcsname[3]%
    {\hbox{\ensuremath{%
          #2{##1}%
          \raisebox{-.1ex}{\ensuremath{\xrightarrow[{\raisebox{.5ex}[0mm][.1ex]{\smash{\tiny #4}}}]%
              {\phantom{.}\raisebox{-0.3ex}[0mm][0.1ex]{{#3{\ensuremath{\scriptstyle ##2}}}}\phantom{.}}}}%
          {}#5{##3}%
        }%
      }}%
  \else%
    \expandafter\newcommand\csname #1sym\endcsname%
    {\ensuremath{\xrightarrow{#3{}}}%
      \index{#4@\protect{\csname #1sym\endcsname}}}%
    \expandafter\newcommand\csname #1name\endcsname%
    {\csname #1sym\endcsname\xspace}%
    \expandafter\newcommand\csname #1\endcsname[3]%
    {\hbox{\ensuremath{%
          #2{##1}%
          \ensuremath{\xrightarrow{#3{##2}}}%
          #5{##3}%
        }%
      }\index{#4@\protect{\csname #1sym\endcsname}}}%
  \fi%
}

\newcommand*{\mkoperel}[4]{%
  \expandafter\newcommand\csname #1sym\endcsname%
       {\raisebox{.2ex}{\ensuremath{\xrightarrow[{\raisebox{.5ex}[0mm][.1ex]{\tiny #3}}]%
             {}}}\index{#3@\protect{\csname #1sym\endcsname}}}%
  \expandafter\newcommand\csname #1name\endcsname%
       {\csname #1sym\endcsname\xspace}%
  \expandafter\newcommand\csname #1\endcsname[2]%
       {\hbox{\ensuremath{%
              #2{##1}%
              \raisebox{.2ex}{\ensuremath{\xrightarrow[{\raisebox{.5ex}[0mm][.1ex]{\tiny #3}}]{}}}%
              #4{##2}%
            }%
          }\index{#3@\protect{\csname #1sym\endcsname}}}%
}

\newcommand*{\mkoperelaa}[2]{%
  \mkoperel{#1}{\encloseinangles}{#2}{\encloseinangles}}                      
\newcommand*{\mkcomrelaa}[2]{%
  \mkcomrel{#1}{\encloseinangles}{\donotenclose}{#2}{\encloseinangles}}   
\newcommand*{\mkoperelnn}[2]{%
  \mkoperel{#1}{\donotenclose}{#2}{\donotenclose}}                            
\newcommand*{\mkcomrelnn}[2]{%
  \mkcomrel{#1}{\donotenclose}{\donotenclose}{#2}{\donotenclose}}         
\newcommand*{\mkoperelan}[2]{%
  \mkoperel{#1}{\encloseinangles}{#2}{\donotenclose}}                         
\newcommand*{\mkcomrelan}[2]{%
  \mkcomrel{#1}{\encloseinangles}{\donotenclose}{#2}{\donotenclose}}      

\newcommand*{\mkdirrelaa}[3]{%
  \mkcomrel{#1}{\donotenclose}{#2}{#3}{\donotenclose}}         

\newcommand{\VSsym}{\index{visualSTATE@\protect{\VSsym}}\textsf{visual\-STATE}}

\newcommand{\STM}{\index{Statemate@\protect{\STM}}\textsc{statemate}\xspace}
\newcommand{\SCsym}{\expandafter\index{SCOPE@\protect{\SCsym}}\textsf{SCOPE}}

\newcommand{\Rhapsodysym}{\index{Rhapsody@\protect{\Rhapsodysym}}\textsc{Rhapsody}}

\newcommand{\inonek}{\hbox{\ensuremath{\in\kern-.5mm\{1..k\}}}}

\mkfuncp{elems}{\wordsp{elems}}
\mkfuncp{dom}{\wordsp{dom}}
\mkfuncp{rng}{\wordsp{rng}}
\mkfuncp{Power}{\mathscr{P}}
\mkfuncp{Multi}{\mathcal{M}}
\mkfuncp{Forest}{\mathcal{F}}

\mkfunc{List}{}{\followwithast}
\newcommand{\Projectsym}[1]{\ensuremath{\pi_{#1}}}
\newcommand{\Project}[2]
        {\ensuremath{\Projectsym{#1}({#2})}\xspace}

\mktype{OR}{or}             
\mktype{XOR}{xor}           
\mktype{AND}{and}    

\mkkeyword{Ids}{\wordsp{Id}}
\mkkeyword{State}{\wordsp{State}}
\mkkeyword{History}{\wordsp{History}}

\newcommand*{\ORst}{\expandafter\index{or-state@\protect{\ORst}}\ORsym{}-state\xspace}
\newcommand*{\ORsts}{\expandafter\index{or-state@\protect{\ORst}}\ORsym{}-states\xspace}
\newcommand*{\XORst}{\expandafter\index{xor-state@\protect{\XORst}}\XORsym{}-state\xspace}
\newcommand*{\XORsts}{\expandafter\index{xor-state@\protect{\XORst}}\XORsym{}-states\xspace}
\newcommand*{\ANDst}{\expandafter\index{and-state@\protect{\ANDst}}\ANDsym{}-state\xspace}
\newcommand*{\ANDsts}{\expandafter\index{and-state@\protect{\ANDst}}\ANDsym{}-states\xspace}

\newcommand*{\StateORsym}{\index{\protect{\StateORsym}}\ensuremath{\State_{\text{\ORsym}}}}   
\newcommand*{\StateANDsym}{\index{\protect{\StateANDsym}}\ensuremath{\State_{\text{\ANDsym}}}}

\mkfuncp{type}{\Gamma_S}

\newcommand{\sbstatekern}{\kern -0.18em}

\newcommand{\sbstatesym}{\ensuremath{\searrow}\xspace}  
\newcommand{\notsbstatesym}{\ensuremath{\not\searrow}\xspace}

\newcommand{\sbstateplsym}{\ensuremath{\sbstatesym\kern -0.66em^{+}\kern 0.13em}}
 
\newcommand{\notsbstatepl}[2]{\hbox{\ensuremath{#2\sbstatekern\notsbstatesym\kern -0.66em^{+}#1}}\xspace} 
\newcommand{\sbstatestsym}{\ensuremath{\sbstatesym\kern -0.66em^\ast\kern 0.13em}}
 
\newcommand{\notsbstatest}[2]{\hbox{\ensuremath{#2\sbstatekern\notsbstatesym\kern -0.66em^\ast#1}}\xspace}

\mkfuncp{priority}{\vartriangleleft}
\mkfuncp{stpriority}{\blacktriangleleft}

\mkfuncp{parent}{\wordsp{parent}}
\mkfuncp{parparent}{\wordsp{parent}^2}
\mkfuncp{children}{\wordsp{children}}
\mkfuncp{ancestors}{\wordsp{ancest}}
\mkfuncp{descendants}{\wordsp{descend}}
\mkfuncp{siblings}{\wordsp{siblings}}

\mkfuncp{ancestorspl}{\wordsp\ancestorssym^{+}}
\mkfuncp{descendantspl}{\wordsp\descendantssym^{+}}
\mkfuncp{ancestorsst}{\wordsp\ancestorssym^\ast}
\mkfuncp{descendantsst}{\wordsp\descendantssym^\ast}
\mkfuncp{siblingsst}{\wordsp{\siblingssym}^\ast}

\mkfuncp{NCA}{\wordsp{NCA}}

\newcommand{\notorthogonalsym}{\hbox{\ensuremath{\not{\kern -2.2pt\bot\kern +2.2pt}}}}

\mkfuncp{initial}{\wordsp{ini}}
\mkfuncp{default}{\wordsp{default}}
\mkfuncp{runhistory}{\eta}

\mkfuncp{entrypathOR}{\entrypathsym_{\text{\ORsym}}}
\mkfuncp{entrypathAND}{\entrypathsym_{\text{\ANDsym}}}
\mkfuncp{entrypath}{\wordsp{entr}}
\mkfuncp{entrysigs}{\wordsp{ensigs}}

\mkfuncp{entrORts}{\entrypathORsym\kern-6pt^{\wordsp{ts}}}
\mkfuncp{entrANDts}{\entrypathANDsym\kern-10pt^{\wordsp{ts~}}}
\mkfuncp{exitpath}{\wordsp{extr}}
\mkfuncp{exitsigs}{\wordsp{exsigs}}

\mkkeyword{Id}{\wordsp{Id}}
\mkkeyword{Var}{\wordsp{Var}}
\mkkeyword{Value}{\wordsp{Value}}

\mkkeyword{ExtVar}{\Var_E}
\mkkeyword{IntVar}{\Var_I}
\mkkeyword{Type}{\wordsp{Type}}
\mkkeyword{SimpleType}{\wordsp{SimpleType}}

\mkkeyword{anatomy}{\wordsp{anatomy}}

\mktype{tint}{int}
\mktype{tuint}{uint}
\mktype{treal}{real}
\mktype{tfloat}{float}
\mktype{tdouble}{double}
\mktype{tvoid}{void}

\mkfuncp{Dom}{D}

\mkfuncp{vartype}{\Gamma_V}
\mkfuncp{Array}{\wordsp{Vector}}
\mkfuncs{toracle}{\tau}

\mkfuncp{mustclosure}{\raisebox{-2pt}{$\Box$}\textit{-closure}}

\mkkeyword{Event}{\wordsp{Event}}
\mkkeyword{Signal}{\wordsp{Signal}} 
\mkkeyword{External}{\wordsp{External}} 
\mkkeyword{Guard}{\wordsp{Guard}}
\mkkeyword{Ebind}{\wordsp{Ebind}}
\mkkeyword{Einst}{\wordsp{Einst}}
\mkkeyword{Sexp}{\wordsp{Sexpr}}
\mkkeyword{Lvalue}{\wordsp{Access}}

\mkfuncp{evtype}{\Gamma_E}

\mkkeyword{topstate}{\wordsp{root}}

\mkkeyword{Exp}{\wordsp{Exp}}
\mkkeyword{OExp}{\wordsp{exp}}
\mkkeyword{Aexp}{\wordsp{Aexp}}
\mkkeyword{Bexp}{\wordsp{Bexp}}

\mkkeyword{Assgn}{\wordsp{Assgn}}
\mkkeyword{Output}{\wordsp{Output}}
\mkkeyword{Oinst}{\wordsp{Oinst}}
\mkkeyword{Ainst}{\wordsp{Ainst}}
\mkkeyword{Oexp}{\wordsp{Oexpr}}

\mkfuncp{fntype}{\Gamma_F}

\mkkeyword{Action}{\wordsp{Action}}
\mkkeyword{Trans}{\wordsp{Trans}}

\mkfuncp{params}{\wordsp{params}}
\mkfuncp{scope}{\wordsp{scope}}
\mkfuncp{iscope}{\wordsp{iscope}}
\mkfuncp{scopes}{\wordsp{scope}}
\mkfuncp{poscond}{\wordsp{pos}}
\mkfuncp{negcond}{\wordsp{neg}}
\mkfuncp{targets}{\wordsp{targets}}
\mkfuncp{source}{\wordsp{source}}
\mkfuncp{expr}{\wordsp{expr}}
\mkfuncp{guard}{\wordsp{guard}}
\mkfuncp{entry}{\wordsp{en}}
\mkfuncp{exit}{\wordsp{ex}}

\mkfuncp{freevars}{\wordsp{fvars}}
\mkfuncp{pure}{\wordsp{pure}}
\mkfuncp{closed}{\wordsp{closed}}

\mkkeyword{Store}{\wordsp{Store}}
\mkkeyword{Queue}{\wordsp{Queue}}
\mkkeyword{state}{\sigma}
\mkkeyword{gstate}{\Sigma}

\mkfuncp{store}{\varrho}
\mkfuncp{history}{\wordsp{his}}

\mkkeyword{queue}{q}

\mkoperelnn{trarrow}{\stignore{0transitionrelation}}
\mkoperelaa{access}{\textrm{access}}
\mkoperelaa{Aeval}{\textrm{Aeval}}
\mkoperelan{Aexec}{\textrm{asgn}}
\mkoperelan{Beval}{\textrm{Beval}}
\mkoperelaa{Cexec}{\textrm{C-sem}}
\mkcomrelnn{comarrow}{}
\mkcomrelnn{Init}{\textrm{init}}
\mkcomrelaa{Macro}{\textrm{macro}}
\mkcomrelaa{Micro}{\textrm{micro}}
\mkoperelaa{Mini}{\textrm{mini}}
\mkcomrelaa{Oeval}{\textrm{output}}
\mkcomrelan{Rexec}{\textrm{exec}}  
\mkcomrelnn{Senter}{\textrm{enter}}
\mkoperelaa{Seval}{\textrm{signal}}
\mkcomrelnn{Sexit}{\textrm{exit}}
\mkcomrelaa{systr}{\stignore{0transitionrelation}}
\mkcomrelnn{Tfire}{\textrm{fire}}

\mkfuncp{cons}{\wordsp{cons}}

\mkfuncn{tail}{\text{\bfseries tl}\,}

\mkfuncn{etail}{{\textbf{src}}\,}
\mkfuncn{ehead}{{\textbf{tgt}}\,}

\mkfuncp{Br}{\textsf{Br}}

\mkfuncp{activeset}{\wordsp{enabled}}
\mkfuncp{enabled}{\wordsp{enabled}}
\mkfuncp{resactive}{\wordsp{resolved-enabled}}
\mkfuncp{activestate}{\wordsp{active}}
\mkfuncp{inactivestate}{\wordsp{inactive}}

\mkfuncp{traces}{\wordsp{traces}}

\mkkeyword{PR}{\wordsp{Pr}}
\mkkeyword{PRP}{\PRsym_{\mathcal{P}}}
\mkkeyword{PRS}{\PRsym_{\ast}}
\mkkeyword{ENV}{\wordsp{Env}}
\mkkeyword{GEN}{\wordsp{Gen}}
\mkkeyword{GENP}{\GENsym_\mathcal{P}}
\mkkeyword{GENS}{\GENsym_\ast}
\mkkeyword{OBS}{\wordsp{Obs}}
\mkkeyword{OBSP}{\OBSsym_\mathcal{P}}
\mkkeyword{OBSS}{{\OBSsym_\ast}}

\mkkeyword{OUT}{\wordsp{Out}}
\mkkeyword{IN}{\wordsp{In}}

\newcommand{\gentrsymname}{}
\newcommand{\obstrsymname}{}
\newcommand{\nottrsymname}{}
\mkdirrelaa{gentr}{\followwithbang}{\gentrsymname}
\mkdirrelaa{obstr}{\followwithqstn}{\obstrsymname}
\mkdirrelaa{nottr}{\followwithnot}{\nottrsymname}

\mkternaryrel{relsim}{\leqslant}
\mkternaryrel{relbisim}{\sim}
\mkternaryrel{twowayrelsim}{\index{simulation!relativized, two way}\lessgtr}
\mkbinaryrel{twowaysim}{\index{simulation!two way}\lessgtr}
\mkbinaryrel{osim}{\leqslant}
\mkbinaryrel{gsim}{\leqslant}
\mkbinaryrel{obisim}{\sim}
\mkbinaryrel{gbisim}{\sim}
\mkbinaryrel{syssim}{\leqslant}
\mkbinaryrel{modref}{\leq_{\textnormal{\textrm{m}}}}
\mkbinaryrel{altsim}{\leq_{\textnormal{\textrm{a}}}}
\mkternaryrelraised{omodref}{\leq^{\ast}_{\textnormal{\textrm{m}}}}
\mkternaryrelraised{agmodref}{\leq^{\ast}_{\textnormal{\textrm{ag}}}}
\mkternaryrelraised{esopref}{\leq^{\triangleleft}_{\textnormal{\textrm{m}}}}
\mkbinaryrel{oaltsim}{\leq^{\ast}_{\textnormal{\textrm{a}}}}
\mkbinaryrel{envsim}{\leqslant}
\mkbinaryrel{sysbisim}{\sim}
\mkbinaryrel{discr}{\index{discrimination}\sqsubseteq}
\mkbinaryrel{implements}{\preceq}
\mkbinaryrel{twodiscr}{\raisebox{1.25ex}{\rotatebox[origin=ct]{-180}%
    {\index{discrimination}\smash{\ensuremath{\sqsupseteq}}}}}
\mkbinaryrel{bidiscr}{\raisebox{-.5ex}{\ensuremath{%
      \index{discrimination}\stackrel{\sqsubset}{\scalebox{0.8}{\ensuremath{\sim}}}}}}

\mkbinaryrel{composetr}{\!\upharpoonright\!}
\mkbinaryrel{crossprod}{\otimes}

\mkkeyword{IOATS}{\text{\index{IOATS}IOATS}}
\mkkeyword{IOATSs}{\text{\index{IOATS}IOATSs}}

\mkfuncp{SSF}{\mathbb{S}}

\mkkeyword{SET}{\textbf{Set}}
\mkkeyword{set}{\textbf{set}}
\mkkeyword{MSET}{\textbf{M-set}}
\mkkeyword{mset}{\textbf{m-set}}
\mkkeyword{INTER}{\textnormal{\textbf{Inter}}}
\mkkeyword{inter}{\textnormal{\textbf{inter}}}

\newcommand{\BLINDsym}{\ensuremath{\index{environment!blind}\mathbf{B}}}

\mkfuncsub{parBLIND}{\BLINDsym}

\newcommand{\UNIVsym}{{\ensuremath{\index{environment!perfect vision}\mathbf{V}}}}

\mkfuncsub{parUNIV}{\UNIVsym}

\mkfuncp{eqdom}{\wordsp{part}}

\newcommand{\WBS}[2][\null]
    {\ensuremath{\textit{WBS}\left(#2,
     \ifthenelse{\equal{#1}{\null}}{\emptyset}{#1}\right)}\xspace}

\usepackage{fixme}
\FXRegisterAuthor{ad}{AD}{Alexandre}
\FXRegisterAuthor{kgl}{KGL}{Kim}
\FXRegisterAuthor{al}{AL}{Axel}
\FXRegisterAuthor{mm}{MM}{Marius}
\FXRegisterAuthor{pb}{PB}{Peter}

\makeatletter
\newcommand{\xRightarrow}[2][]{%
     \ext@arrow 0055{\Rightarrowfill@}{#1}{#2}%
} 
\makeatother

\renewcommand{\phi}{\varphi}

\newcommand{\uppaal}{\textsc{Uppaal}\xspace}

\newcommand{\uppaalsmc}{\textsc{Uppaal-smc}\xspace}

\newcommand{\forget}[1]{}

\def\titlerunning{Distributed Parametric and Statistical Model Checking}
\def\authorrunning{Bulychev, David, Larsen, Legay, Mikučionis}

\begin{document}
\initfloatingfigs

%
%

\title{
Distributed Parametric and Statistical Model Checking
\thanks{
Work partially supported by VKR Centre of Excellence -- MT-LAB,
an ``Action de Recherche Collaborative'' ARC (TP)I, and the CREATIVE project ESTASE.
}
}

\author{
Peter~Bulychev\quad
Alexandre~David\quad
Kim~Guldstrand~Larsen\quad
Marius~Mikučionis\quad
\institute{Department of Computer Science\\
Aalborg University, Denmark}
\email{\{pbulychev,adavid,kgl,marius\}@cs.aau.dk}
\and
Axel~Legay
\institute{INRIA Rennes, France\\
Department of Computer Science\\
Aalborg University, Denmark}
\email{alegay@irisa.fr}
}

\maketitle

\begin{abstract}
Statistical Model Checking (SMC) is a trade-off between testing and
formal verification. The core idea of the approach is to conduct some
simulations of the system and verify if they satisfy some given
property. In this paper we show that SMC is easily parallelizable on a
master/slaves architecture by introducing a series of algorithms that
scale almost linearly with respect to the number of slave
computers. Our approach has been implemented in the UPPAAL SMC toolset
and applied on non-trivial case studies.
\end{abstract}

\section{Introduction}

Computers play a central role in modern societies and their errors can
have dramatic consequences. For example, such errors could jeopardize
a banking system, possibly stalling the economy of a whole country or,
more dramatically, endanger human life through the failure of some
safety critical systems (railway signaling, integrated avionics,
air-traffic, medical life support machines, automotive
electronics). It is therefore not surprising that proving the
correctness of computer systems is a highly relevant
problem. Unfortunately, the growing complexity in system design makes
it almost impossible to ensure correctness simply by looking at the
(possibly distributed) code. Automatic techniques are thus needed.

The most common method to ensure the correctness of a system is {\em
  testing} (see \cite{BJKLP05} for a survey). After the computer
system is constructed, it is tested using a number of {\em test cases}
with predicted outcomes. Testing techniques have shown effectiveness
in bug hunting in many industrial problems. Unfortunately, testing is
not always the perfect solution. Indeed, since there is, in general,
no way for a finite set of test cases to cover all possible scenarios,
errors may remain undetected. There are also methods that can ensure
the full correctness of a system. Those methods, also called {\em
  formal methods}, use mathematical techniques to check whether the
system will behave correctly for all possible scenarios. Over the
past, formal methods such as {\em symbolic model
  checking}\,\cite{McMillan93} have been used to verify systems with
more than $10^{20}$ reachable states\,\cite{BCMDH92}.

In an ideal world, it would thus be ``better'' to use formal methods
rather than testing ones. Unfortunately, improvements in development
of formal methods do not seem to follow the increasing complexity in
system design. Nowadays, most of formal methods suffer from the
so-called {\em state-space explosion problem}, which makes them
unenforceable to large industrial case studies. As testing does not
suffer from the same problem, it remains the only scalable technique
and it is thus the one promoted by the industrials.

As  we already  said,  the major  drawback  with testing  is that,  in
general, it  does not  give any confidence  on the correctness  of the
entire system. This lack of  accuracy has motivated the development of
new  algorithms  that  combine  testing  techniques  with  statistical
algorithms.   These  techniques, also  called  {\em Statistical  Model
  Checking techniques}  (SMC)\,\cite{HLMP04,SVA04,You05a}, can be seen
as a trade-off between testing  and formal verification. The core idea
of  the approach  is to  conduct some  simulations of  the  system and
verify if they satisfy some  given property. The results are then used
together with algorithms from the  statistical area in order to decide
whether    the    system   satisfies    the    property   with    some
probability. Statistical model checking techniques can also be used to
estimate   the   probability  that   a   system   satisfies  a   given
property\,\cite{HLMP04,GS05}.  Of course, in contrast to an exhaustive
approach,  a simulation-based  solution does  not guarantee  a correct
result with  100\% confidence.  However, it is  possible to  bound the
probability of making an  error. Simulation-based methods are known to
be far  less memory and time  intensive than exhaustive  ones, and are
sometimes the  only option\,\cite{YKNP06,JKOSZ07}. Among  existing SMC
algorithms, one find those that use a fixed number of samplings (those
to  estimate  the  probability)  and  those  that  support  sequential
sampling (those that  test an estimate of the  probability provided by
the  user)   where  the   number  of  simulation   is  not   known  in
advance\,\cite{Wal04}.

Statistical model  checking gets  widely accepted in  various research
areas  such  as software  engineering,  in  particular for  industrial
applications, or  even for  solving problems originating  from systems
biology\,\cite{CFLHJL08,JCLLPZ09}. There are  several reasons for this
success.  First, SMC is very simple to understand, implement, and use.
Second, it  does not require  extra modeling or  specification effort,
but simply a stochastic operational  sematics of the model that can be
used  as the  basis  for simulation  and  checked against  state-based
properties.        Third,        it       allows       to       verify
properties\,\cite{CDL08,BBBCDL10}   that   cannot   be  expressed   in
classical temporal logics.

However, SMC  is not a panacea  and many huge size  problems are still
beyond  its scope.  Indeed, sometimes  the  algorithm needs  a lot  of
simulation to compute,  and the computation of each  simulation may be
time consuming.  There are  two solutions to  this problem.  The first
solution is  to propose  new algorithms and  heuristics to  reduce the
number   of  simulations  needed   for  the   algorithm  to   reach  a
decision. The second approach consists in taking new and emerging
platforms into account. This paper goes  for the second
solution. A  trend to speed up  computation time and  hence to improve
the  efficiency  of SMC  is  certainly to  take  advantage  of the  new
technologies  among  which  one  find  our ability  to  use  several
computers  working  in  parallel.  In  fact,  it  is  well-known  that
statistical  solutions   methods  that  use   samples  of  independent
observations  are  often trivially  parallelizable  (see  the work  on
Metropolis and  Ulam). As observed  by Youness, SMC algorithms  can be
distributed  through the  help  of a  master/slave architecture  where
multiple computers are  used to generate the simulations.  The idea is
as  follows: one  or more  slave processes  register their  ability to
generate  simulation with  a single  master  process that  is used  to
collect those simulations and  peform the statistical test. As pointed
out by Youness\,\cite{You05c}, in order to ensure that simulations are
independent, some care needs  to be taken when generating pseudorandom
number on each  machines, which can easily be  solved by incorporating
the   number  of   each  processor   in  the   generation   of  theses
numbers\,\cite{You05c}. When  using sequential testing,  the situation
becomes  more  complex  as  it  is important  to  guarantee  that  the
technique will  not introduce a  bias against simulations that  take a
longer time  to generate.  The latter  can be done  by computing  an a
priori   to  the   order   in  which   simulations   are  taken   into
account. Defining this order so  that the algorithm scales up linearly
with the number of slave processors may be complex and remains a major
challenge through distributing sequential algorithms.

In this paper, we report on the implementation of a new methodology we
use to parallelize the statistical model checking algorithms we
developed for model checking stochastic timed
automata\,\cite{uppaal_formats,uppaal_cav} against weighted temporal
logic properties. Those SMC algorithms, which have been implemented in
\uppaalsmc -- a SMC extension of the \uppaal toolset\,\cite{UPPAAL} --
rely on Wald's sequential hypothesis testing (used to test a
probability) and Monte Carlo simulation (used to estimate a
probability). Our approach, also implemented in \uppaalsmc, scales
better than the one of Youness. Moreover, we show how to perform
parameter estimation with SMC. The latter approach can be used to
optimize a given algorithm (what is the best network topology, the
best transmission rate, ...)  in an efficient manner. Our approach is
applied to non-trivial case studies.

%

\newcommand{\nplus}{\ensuremath{\bbbn_{\geq0}}}
\newcommand{\rplus}{\bbbr_{\geq0}}
\newcommand{\U}{{\cal U}}
\renewcommand{\L}{{\cal L}}
\newcommand{\arrow}[1]{\stackrel{#1}{\longrightarrow}}

\renewcommand{\P}{\ensuremath{{\mathbb P}}\xspace}

\section{Statistical Model Checking}

\subsection{The model}

In this section, we briefly recap the concept of Priced Timed Automata
(PTA), see \cite{uppaal_formats} for more details. We denote
${\cal{B}}(X)$ to be a finite conjunction of bounds of the form $x\sim
n$ where $x\in X$, $n\in\bbbn$, and $\sim\in\{<,\leq, >, \geq \}$. A
{\em Priced Timed Automaton} (PTA) is a tuple ${\cal
  A}=(L,\ell_0,X,E,R,I)$ where: (i) $L$ is a finite set of locations,
(ii) $\ell_0\in L$ is the initial location, (iii) $X$ is a finite set
of real-valued variables called clocks, (iv) $E\subseteq L \times
{\cal{B}}(X) \times 2^X \times L$ is a finite set of edges, $(v)$
$R:L\rightarrow\mathbb{Z}_{\geq 0}$ assigns a rate vector to each
location, and (vi) $I:L\rightarrow{\cal{B}}(X)$ assigns an invariant
to each location. A state of a PTA is a pair $(l,v)$ that consists of
a location $l$ and a valuation of clocks $\nu : X \rightarrow \rplus$.
From a state $(l,v) \in L \times \rplus^X$ a PTA can either let time
progress or do a discrete transition and reach a new location.  During
time delay clocks are growing with the rates defined by $R(l)$, and
the resulting clock valuation should satisfy invariant $I(l)$.  A
discrete transition from $(l, v)$ to $(l', v')$ is possible if there
is $(l, g, Y, l') \in E$ such that $v$ satisfies $g$ and $v'$ is
obtained from $v$ by resetting clocks from the set $Y$ to $0$.  A run
of PTA is a sequence of alternating time and discrete transitions.

Several PTA $M_1, M_2, \dots, M_n$, can be put in parallel via message
passing in order to form a network $M_1 \| M_2 \| \dots \| M_n$ of
PTAs. By message passing, we mean that the automata can synchronize on
some transitions and exchange messages through input and output
actions.

In  order to  perform  SMC  on PTAs,  we  have to  equip  them with  a
stochastic semantic.  The latter being needed to  define a probability
space  over  the sets  of  their  executions.  Giving details  on  the
stochastic  semantic of PTAs  is beyond  the scope  of this  paper but
details are available  in \cite{uppaal_formats}. Roughly speaking, the
stochastic semantic  associates probability distributions  on both the
delays  one can spend  in a  given state  as well  as on  a transition
between  states. In  general  one considers  uniform distribution  for
bounded  delays and  exponential for  the case  where a  component can
remain indefinitely in a  state. As observed in \cite{uppaal_formats},
though the stochastic semantic of each individual PTA is rather simple
(but quite realistic), arbitrarily  complex stochastic behavior can be
obtained  by their  composition when  mixing  individual distributions
through  message  passing. The  beauty  of  our  model is  that  these
distributions are  naturally and automatically defined  by the network
of PTAs.

Our implementation supports extensions of PTA, coming from the language
of the \uppaal model checker~\cite{UPPAAL}.  Such models can contain
integer variables that can be present in transition guards, and they
can be updated only when a discrete transition is taken.
Additionally, we support other features of the \uppaal model checker's
input language such as data structures and user-defined functions.

A parametrized PTA $M(p)$ is a PTA in which some integer constant
(transition weight or constant in variable assignment/clock invariant)
is replaced by a parameter $p$.

For defining properties we use weighted temporal logic PWCTL, which
contains formulas of the form $\Diamond_{c\leq C}\phi$.  Here $c$ is
an observer clock (that is never reset and should grow to infinity on
any infinite run of PTA), $C \in \rplus$ and $\phi$ is a
state-predicate.  We say that a run $\pi$ satisfies
$\psi=\Diamond_{c\leq C}\phi$ if there exists $(l, v) \in \pi$ such
that $l$ satisfies $\phi$ and $v(c)\leq C$.  We denote by $Pr_{\cal A}[\psi]$ the probability that a random run of the model ${\cal A}$
satisfies $\psi$.

\subsection{Statistical Model Checking for NPTAs}

The problem  of checking   $Pr_{\cal A}[\Diamond_{c\leq C}\phi]\geq p$
(${\cal A}$ being  a PTA and $p\in[0,1]$) is unfortunately undecidable in general \footnote{Exceptions being PTA with 0 or 1 clocks.}. Our solution  is to
approximate the  answer using simulation-based  algorithms known under
the name  of statistical model  checking algorithms. We  briefly recap
statistical algorithms permitting to answer the following two types of
questions :
\begin{enumerate}
\item  {\sl  Testing:}  Is  the  probability  $Pr_{\cal
    A}[\Diamond_{c\leq  C}\phi]$ for  a given  NPTA $\boldsymbol{{\cal
      A}}$ greater or equal to a certain threshold $\theta$ ?
\item
{\sl Estimation:} What is the probability $Pr_{\cal
  A}[\Diamond_{c\leq  C}\phi]$  for a given NPTA $\boldsymbol{{\cal
    A}}$?
\end{enumerate}

From a conceptual point of view both solving the two above questions
via SMC is simple. First, each run of the system is encoded as a
Bernoulli random variable that is true if the run satisfies the
property and false otherwise. Then a statistical algorithm groups the
observations to answer the two questions. For the qualitative
question, we shall use sequential hypothesis testing, while for the
quantitative question we will use an estimation algorithm that
ressemble the classical Monte Carlo simulation. The two solutions are
detailed hereafter.

\paragraph{\bf  Sequential Sampling/Testing} This approach reduces the qualitative question to the test the
hypothesis $H: p=\P_{\boldsymbol{{\cal A}}}(\Diamond_{C\le c}\phi) \ge
\theta$ against $K: p < \theta$. To bound the probability of making
errors, we use strength parameters $\alpha$ and $\beta$ and we test
the hypothesis $H_0: p \ge p_0$ and $H_1: p \le p_1$ with
$p_0=\theta+\delta_0$ and $p_1=\theta-\delta_1$. The interval
$p_0-p_1$ defines an indifference region, and $p_0$ and $p_1$ are used
as thresholds in the algorithm. The parameter $\alpha$ is the
probability of accepting $H_0$ when $H_1$ holds (false positives) and
the parameter $\beta$ is the probability of accepting $H_1$ when $H_0$
holds (false negatives). The above test can be solved by using Wald's
{\em sequential hypothesis testing}\,\cite{Wal04}.  This test computes
a proportion $r$ among those runs that satisfy the property. With
probability 1, the value of the proportion will eventually cross
$\log(\beta/(1-\alpha)$ or $\log((1-\beta)/\alpha)$ and one of the two
hypothesis will be selected.

\paragraph{\bf Estimation}
This algorithm~\cite{HLMP04} computes the number of runs needed in
order to produce an approximation interval
${\lbrack}p-\epsilon,p+\epsilon{\rbrack}$ for $p=Pr(\psi)$ with a
confidence $1-\alpha$. The values of $\epsilon$ and $\alpha$ are
chosen by the user and the number of runs relies on the
Chernoff-Hoeffding bound.

\section{Distributed Statistical Model-Checking}

We  report on preliminary  results on  using distributed  computing to
speed-up  SMC algorithms.  We  start by  discussing  the solution  for
hypothesis testing where the number  of simulations needed by the test
is  not  known  in  advance.  A naive  solution  in  distributing  the
generation of the  runs may give rise to a  \emph{bias} in the result,
as  pointed  by Younes~\cite{You05a}.  In  short,  some computers  may
generate (for  example) positive answers more quickly  than some other
computers,   which  may   bias  the   decision  toward   the  positive
answer.  This would not  happen when  computing runs  sequentially. In
general, the time required to generate runs may not be uniform and can
cause  this  type  of  bias.  To  counter  this,  Younes~\cite{You05a}
proposed  a  round-Robin  solution  where  the  runs  are  counted  in
rounds. To  improve performance, Younes  defined safe lower  and upper
bounds on the Binomial random  variable that represents the sum of all
the positive  realisations, i.e., all  the simulation that  do satisfy
the property. Instead of waiting for  the results of all the nodes, if
a result is missing the lower and upper bounds are used to take a safe
decision. This  has the potential  to reduce the execution  time since
decisions may be taken earlier.

\begin{figure}[htb]
\centering
\includegraphics[width=0.3\linewidth]{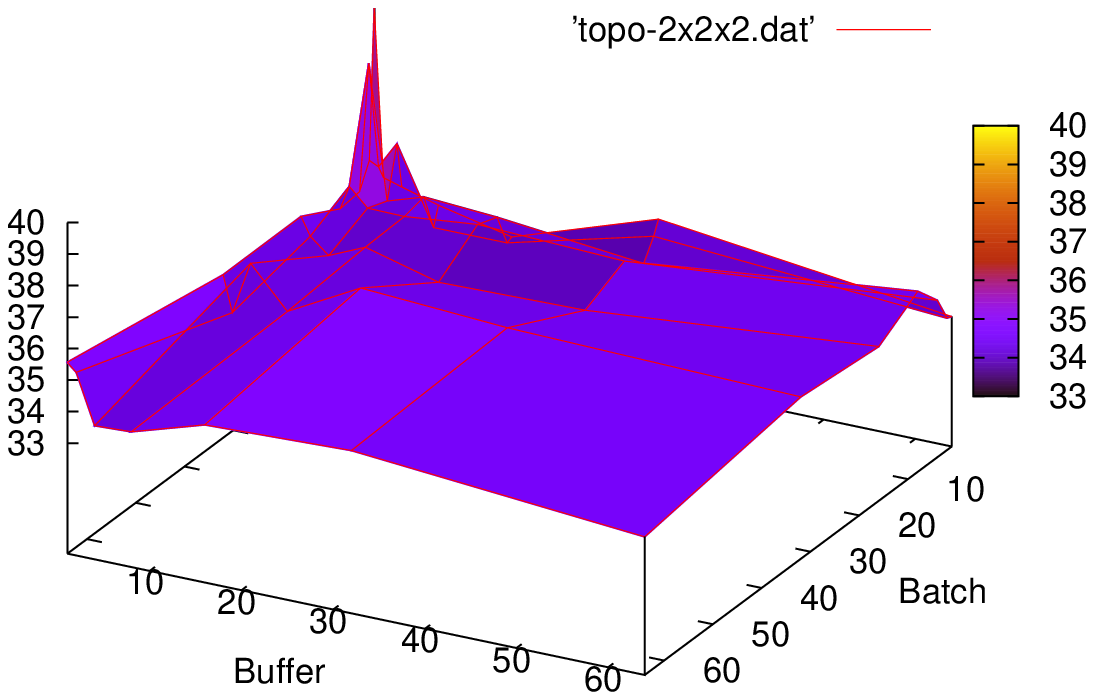}
\includegraphics[width=0.3\linewidth]{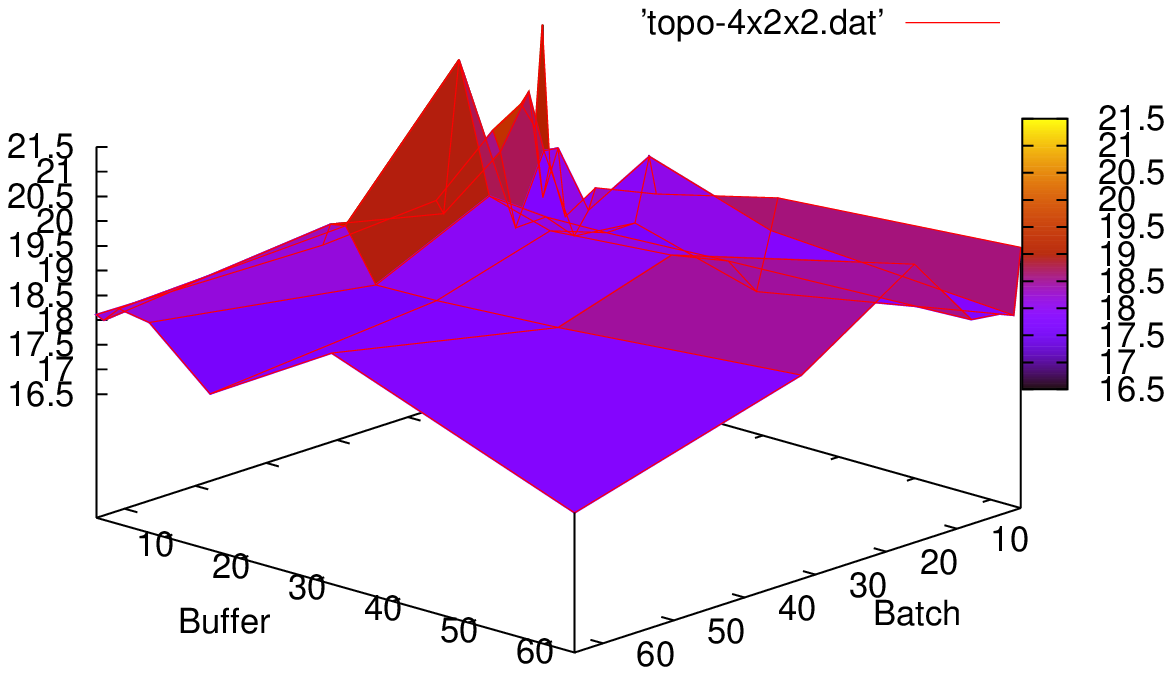}
\includegraphics[width=0.3\linewidth]{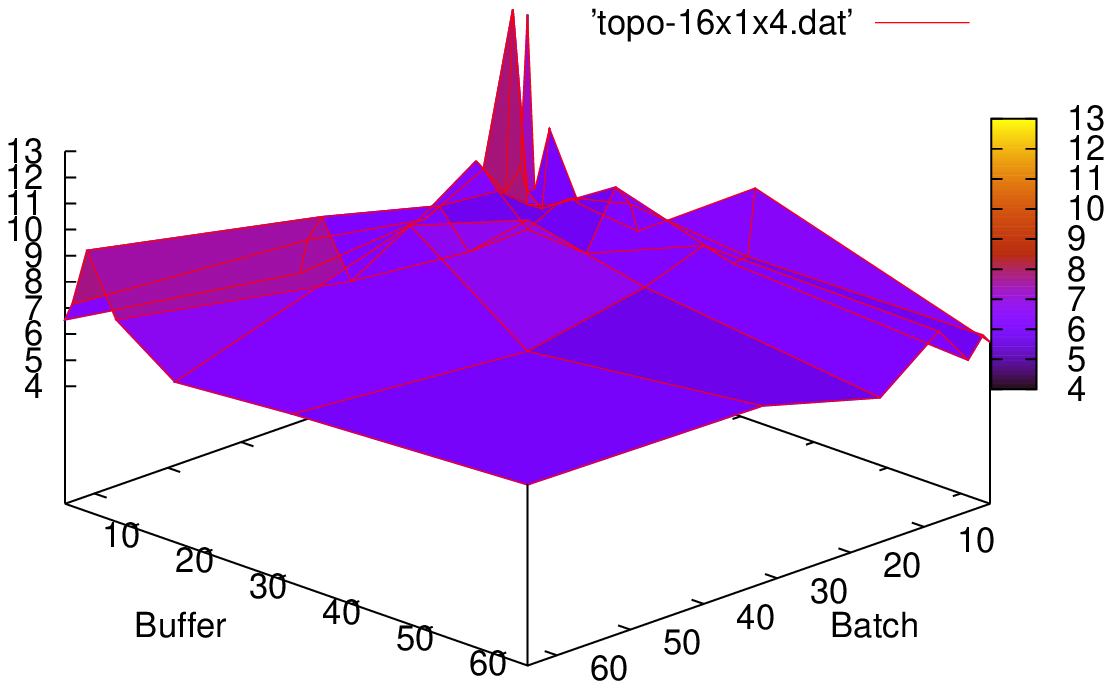}
\caption{Time to check for mutual exclusion for 20 trains
  qualitatively.}
\label{fyrkat}
\end{figure}

We generalize Younes' algorithm by sending the result of simulations
by batches and also by implementing a buffer of incoming result. The
batch is used to reduce communication by sending an aggregate result
of predefined size (instead of individual results). The buffer is used
to improve concurrency since the nodes are more loosely
synchronized. We experiment on these two dimensions for different
topologies, while Younes' algorithm is the particular case where both
are equal to one, which is not very scalable since this generates a
lot of traffic and the nodes are more synchronized.
Figure~\ref{fyrkat} shows the time it took to verify that the mutual
exclusion property of the \emph{train-gate} example distributed with
\uppaal holds with probability 98\% configured with 20 trains and
99.999\% confidence. We show the results for different topologies of
our cluster, NxPxC where N is the number of nodes, P the number of
processors per node, and C the number of cores per processor.

We see, modulo experimental variations\footnote{Clusters are shared
  resources with varying load so results are expected to vary.}, that
the algorithm improves when the batches or buffer are increased but
then it becomes quickly insensitive to these parameters.

Distributing the estimation algorithm is much simpler. We need a
\emph{fixed} number of runs determined by the Chernoff's
bound~\cite{HLMP04} to conclude on a probability value with given
confidence level. This is an embarrassingly parallel problem since we
can simply divide the work equally and gather the result at the
end. To compensate for fluctuations in the cluster, we could implement
work-stealing but as our experiments show, this does not seem to be
necessary since the observed performance scales almost linearly.
 The loss in efficiency in the later cases exhibits the
overhead of starting up all the processors (around 3-4 seconds), which
would be compensated for much longer runs. Figure~\ref{firewire} shows
running time and relative efficiency for checking a few quantitative
properties on the Firewire and LMAC protocol\footnote{The model and
  properties are available on
  \href{http://people.cs.aau.dk/~adavid/smc/}{http://people.cs.aau.dk/\~{}adavid/smc/}.}.

\begin{table}[htb]
\centering
\begin{small}
\begin{tabular}{*{1}{r}|*{5}{r}|*{5}{r}}
\toprule
\multicolumn{1}{l|}{} &
\multicolumn{5}{c|}{Firewire} &
\multicolumn{5}{c}{LMAC} \\
PxC/N  & 1 & 2 & 4 & 8 & 16
            & 1 & 2 & 4 & 8 & 16  \\ 
\hline
\hline
1x1 & 621.7s & 316.7s & 160.2s & 81.1s & 44.7s
       & 279.3s & 140.7s & 73.0s & 37.0s & 19.5s \\
  &\emph{1.00}&\emph{0.98} &\emph{0.97} &\emph{0.96}&\emph{0.87}
  &\emph{1.00}&\emph{0.99}&\emph{0.96}&\emph{0.94}&\emph{0.90}
 \\ \hline
1x2 & 300.9s & 162.2s & 80.5s & 47.6s & 24.3s&
          144.3s & 71.0& 37.5s & 19.2s & 10.4s \\
  &\emph{1.03}&\emph{0.96}&\emph{0.97}&\emph{0.82}&\emph{0.80}
  &\emph{0.97}&\emph{0.98}&\emph{0.93}&\emph{0.91}&\emph{0.84}
 \\ \hline
1x4 & 161.2s & 84.0s & 44.8s & 24.1s & 16.0s&
          74.2s &36.1s&19.3s&9.6s&8.1s \\
  &\emph{0.96}&\emph{0.93}&\emph{0.87}&\emph{0.81}&\emph{0.61}
  &\emph{0.94}&\emph{0.97}&\emph{0.90}&\emph{0.91}&\emph{0.54}
 \\ \hline
2x4 & 85.1s & 46.5s & 23.1s & 14.1s & 8.5&
          35.5s&19.6s&10.1s&10.2s&6.4s\\
  &\emph{0.91}&\emph{0.84}&\emph{0.84}&\emph{0.69}&\emph{0.57}
  &\emph{0.98}&\emph{0.89}&\emph{0.86}&\emph{0.43}&\emph{0.34}
 \\ 
\bottomrule
\end{tabular}
\end{small}
\caption{Time in seconds and efficiency (italic) to checking
  quantitative properties on the Firewire and LMAC model in function of
  the number of nodes (N), processors per node (P) and cores per
  processor (C).}
\label{firewire}
\end{table}

\section{Distributed Parametric Model-Checking: The Principle}

In many practical cases system behaviors depend on the values of a
finite set of constant parameters.  For instance, these parameters can
define network topology, or transmission rate of a node.

An interesting question might be to study how a system behavior
depends on the values of these parameters.  This may include
visualisation of this dependency (drawing plots), optimization/worst
case analysis and determining the correlation between different
parameters.  Another example that we will study below is computing
Nash Equilibrium in wireless ad-hoc networks, e.g. choosing a network
configuration that is stable with respect to the behavior of selfish
nodes.

Let us assume that there is a finite set of parameters, each defined
on a finite domain.  We will model parameterized systems using \uppaal
models in which some integer constants (transition weights or
constants in variable assignment/clock invariants) are parameterized,
e.g. they are replaced by special syntactic constructs that define
the sets of possible values. Currently, we support two constructs:
\begin{itemize}
\item \verb|#range(a, b)| defines the set of all integers between \verb|a| and \verb|b|,
\item \verb|#booleanmatrix(N)| defines the set of all boolean matrices of size \verb|N|, this construct can be used to represent the set of all possible topologies of a network with \verb|N| nodes.
\end{itemize}

We developed a framework for solving the ``parametric'' problems
listed above (visualisation, optimization/worst case analysis, Nash
Equilibrium computation).  In order to solve all these problems our
implementation performs a series of invocations of \uppaalsmc for
different values of parameters.  These invocations are independent of
each other, thus they can be easily distributed on highly
heterogeneous clusters.  Our implementation uses the SLURM batch
system~\cite{SLURM}, or it can submit jobs to the computational nodes
using SSH connection by its own.

\section{Distributed Parametric Model Checking: Case-Studies}

\subsection{Traingate example}

\begin{figure}[!htb]
  \centering
  \begin{tabular}{cc}
    \includegraphics[height=0.28\textheight]{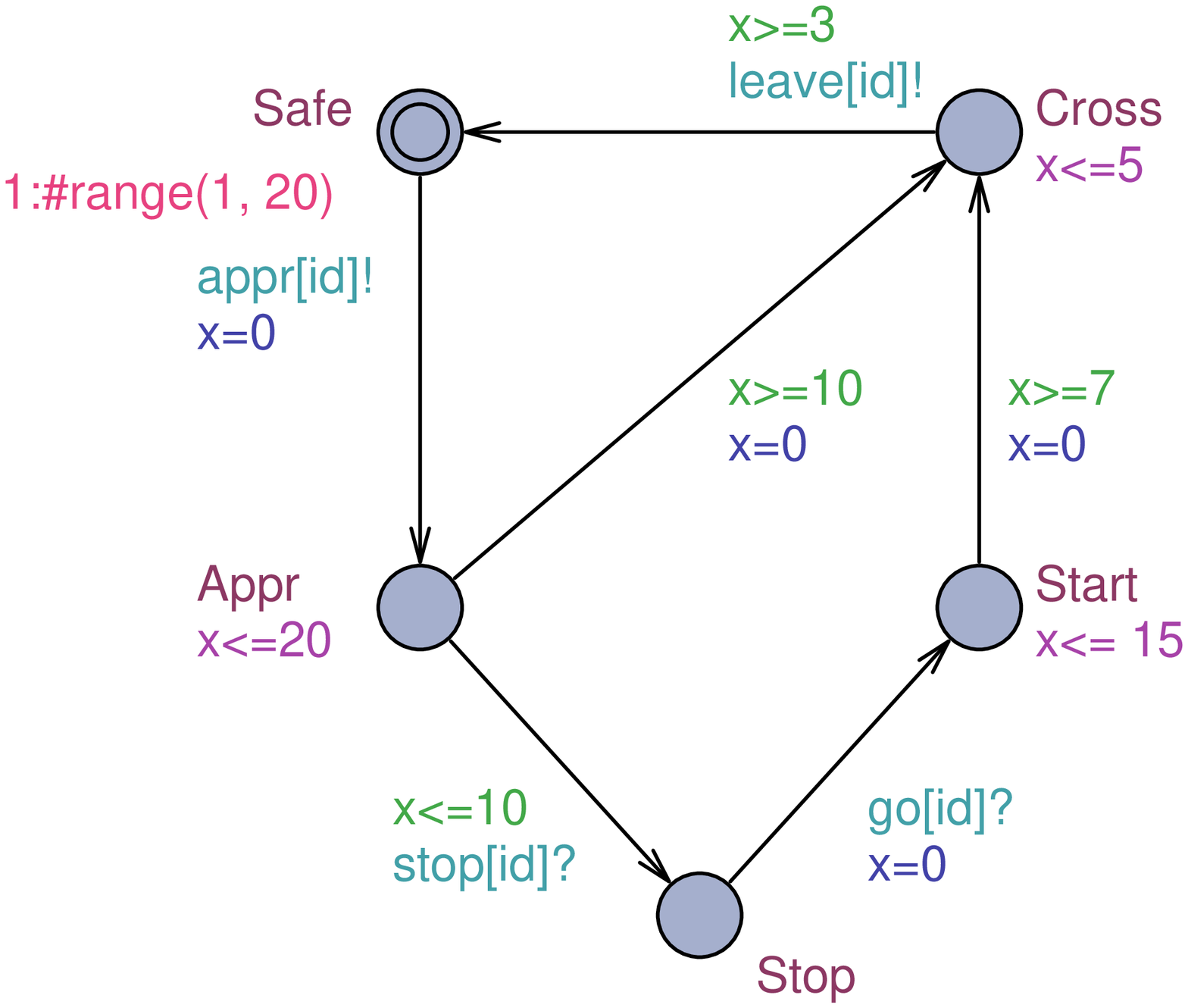} & \includegraphics[height=0.28\textheight]{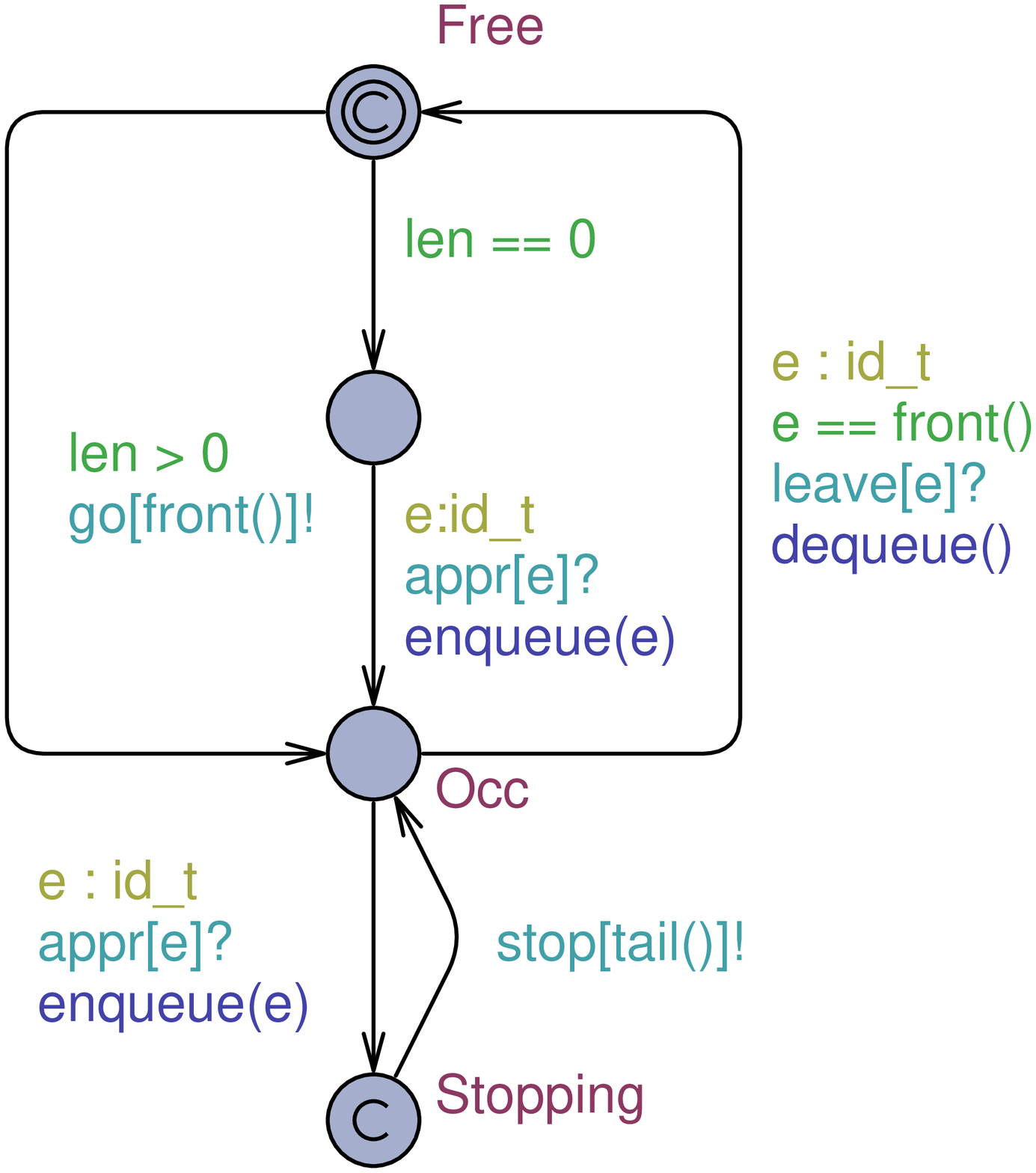}
  \end{tabular}
  \caption{{\sc Uppaal} models of a train (left) and a gate controller (right).}
  \label{fig:traingate_model}
\end{figure}

We consider a model of a railway bridge~\cite{YPD94} where several trains
are crossing a bridge with one track.  Our \uppaal model is depicted on
Fig.~\ref{fig:traingate_model}. Trains start in the {\tt Safe} initial
location where they are not approaching. They will leave that location
and be approaching (and go to location {\tt Appr}) with an arrival rate given by
the expression {\tt 1:\#range(1,20)} on the figure. This is a
parameter declaration that will be used to generate models with values
{\tt 1:1, 1:2, \dots 1:20}. This expression (of the form $i:j$) is an
extension of \uppaal and defines an exponential distribution with the
rate $\frac{i}{j}$ to pick the delay from. When a train is
approaching, it enters {\tt Appr} and synchronizes with {\tt
  appr[id]!}. The gate controller will know that train {\tt id} is
approaching. After 10 time units the train will be crossing (enter
location {\tt Cross}, unless it is stopped before by the gate
controller. This is done with the synchronization {\tt stop[id]?} and
the train goes to the location {\tt Stop}. From there, it is restarted
with the synchronization {\tt go[id]?} by the gate controller and
after 7 time units it will be crossing. After crossing, trains leave
the bridge with {\tt leave[id]!} and are safe again and can decide to approach again.

The gate controller keeps track of stopped trains with a FIFO queue
(not depicted here) that we will not detail. Trains are queued and
dequeued with this queue with the help of functions as seen on the
figure. The gate has two main states {\tt Free} and {\tt Occ}
(i.e. occupied) that keeps track of the state of the bridge. If trains
are approaching then it either stops them if the bridge is occupied or
let them pass otherwise. When the bridge becomes free (one train
leaves), the controller decides to restart a train at the front of the
queue with {\tt go[front()]!}.

Here a (qualitative) safety means to ensure that at most one train can be in the crossing at the same time, and such property can be checked using classical \uppaal model checker.
On top of that, now \uppaalsmc can also evaluate probabilistic (quantitative) properties.
For instance, we can estimate the probability that the first train will cross the bridge within $50$ time units by checking a PWCTL property $\Diamond_{time \leq 50}\big(Train(0).Cross\big)$.

Consider two parameters in our model: the number of trains, and the rate with which these trains are coming. 
The rate parameter is on location `{\tt Safe} shown in Fig.~\ref{fig:traingate_model}, and the number of trains is declared similarly in the \verb|System| declarations.

\begin{figure}[!htb]
  \centering
  \includegraphics[scale=0.6]{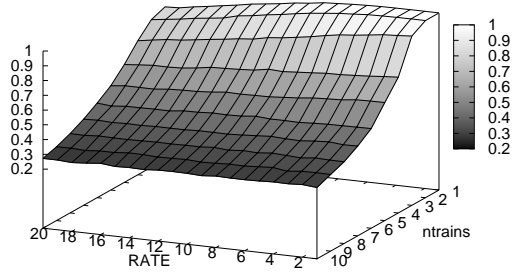}
  \caption{Parametric sweep for the traingate model.}
  \label{fig:train_sweep}
\end{figure}

Fig.~\ref{fig:train_sweep} depicts the results of a parameter sweep of this model. 
The plot shows that when the number of trains increases, the probability that the first
train will cross the bridge within $50$ time units decreases. Indeed, it is more
likely that it will be stopped by other trains (there are more) and
spend time in the \verb|Stop| location. When the arrival rate is
decreased, the probability also decreases.

\subsection{Nash equilibrium Aloha CSMA/CD protocol}
\label{sec:nash}
  Aloha protocol~\cite{aloha} is a simple Carrier Sense Multiple
  Access with Collision Detection (CSMA/CD) protocol that was used in
  the first known wireless data network developed at the University of
  Hawaii in 1971.  The protocol assumes that there are several nodes
  that share the same wireless medium.  Each node is listening to its
  own signal during its transmission and checks that the signal is not
  corrupted by a simultaneous transmission by another node.  In case
  of collision both nodes stop transmitting immediately and wait for a
  random time before they try to transmit again.

\begin{figure}[!htb]
  \centering
  \includegraphics[scale=0.4]{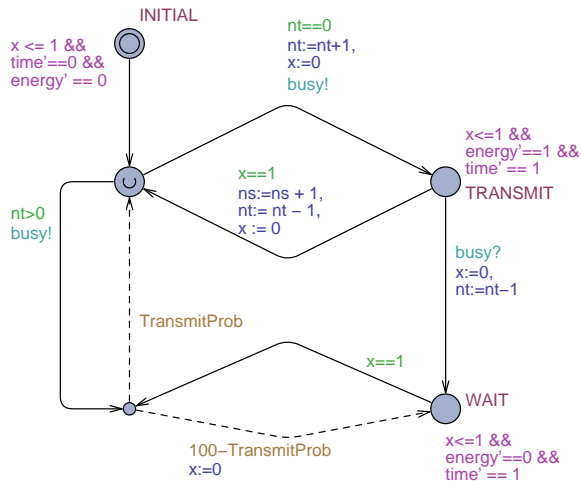}
  \caption{Model of Aloha in \uppaal}
  \label{fig:aloha}
\end{figure}

    The \uppaal model of a single node is given in
    Fig.~\ref{fig:aloha}.  We consider unslotted Aloha where the nodes
    are not necessary synchronized.  Additionally, we study the
    p-persistent variant of Aloha, i.e. a protocol implementation in
    which a random delay before retransmission is distributed
    according to a geometric distribution.  This means that in each
    time slot a node transmits with probability \verb|TransmitProb|
    and waits for one more slot (and then decide again) with
    probability $1-$\verb|TransmitProb|.

  In our experiments we assumed that the goal of a node is to transmit
  a single frame within $50$ time units and to limit energy
  consumption by $3$.  This goal for a node $i$ can be expressed using
  the following PWCTL formula:
  \begin{equation}
    \psi_i \equiv \Diamond_{Node(i).time \leq 50}(Node(i).ns \geq 1 \wedge Node(i).energy \leq 3)
  \end{equation}

  Then the utility function $U_i$ of a node $i$ is equal to the
  probability that the goal $\psi_i$ is satisfied by a random run of a
  system, i.e:
  \begin{equation}
    U_i(p_1, p_2, \dots, p_N) \equiv Pr[S(p_1, p_2, \dots, p_N) \models \psi_i]
  \end{equation}
  , where $p_j$ is equal to the value of \verb|TransmitProb| chosen by node $j$. 

  We consider the case where there is a master node that knows the
  network configuration (here the number of nodes) and broadcasts the
  value of \verb|TransmitProb| parameter to all the nodes.  Now, if there
  are selfish nodes, they can change their values of \verb|TransmitProb| to
  achieve better performance (and other nodes will suffer from that).
  Thus, the interesting question is to find the value of
  \verb|TransmitProb| that satisfies Nash Equilibrium (NE). For such a
  value, it is not profitable for any node to alter its behaviour to
  the detriment of other nodes.  For our case the network is
  symmetric, thus we can search for $NE$ from the point of view of the
  first node only.  In other words, parameter $p$ satisfies NE, iff
  $U_0(p, p, \dots, p)$ is larger than $U_0(p', p, \dots, p)$ for any
  $p'$.

%
%

\begin{table}[!htb]
  \centering
  \caption{Nash equilibrium (NE) and Symmetric optimal (Opt) strategies for Aloha.}
  \label{aloha:results}
 \begin{tabular}{@{\extracolsep{+5pt}}lccccccccc}
\toprule
    Number of nodes\                 		& 2    	& 3    	& 4    	& 5    	&  6    & 7   	\\
\hline 
    NE strategy $p_{NE}$\    		& 0.37 	& 0.40 	& 0.35 	& 0.35 	& 0.41	& 0.42	\\    
    $\widetilde{U}(p_{NE}, p_{NE})$ 				\ 	& 0.99  & 0.98  & 0.95  & 0.89	& 0.75	& 0.61  \\
    Symmetric optimal strategy $p_{opt}$\               			& 0.30 	& 0.30 	& 0.26 	& 0.22	& 0.19	& 0.15 	\\
    $\widetilde{U}(p_{Opt}, p_{opt})$\ 	& 0.99 	& 0.98  & 0.96 	& 0.90	& 0.87  & 0.98  \\
    Computation time \ 					& 2m5s 	& 3m44s & 7m62s & 15m45s& 26m11s& 37m55s  \\
\bottomrule
 \end{tabular}
\end{table}

\begin{figure}[!htb]
  \centering
  \includegraphics[height=0.25\textheight]{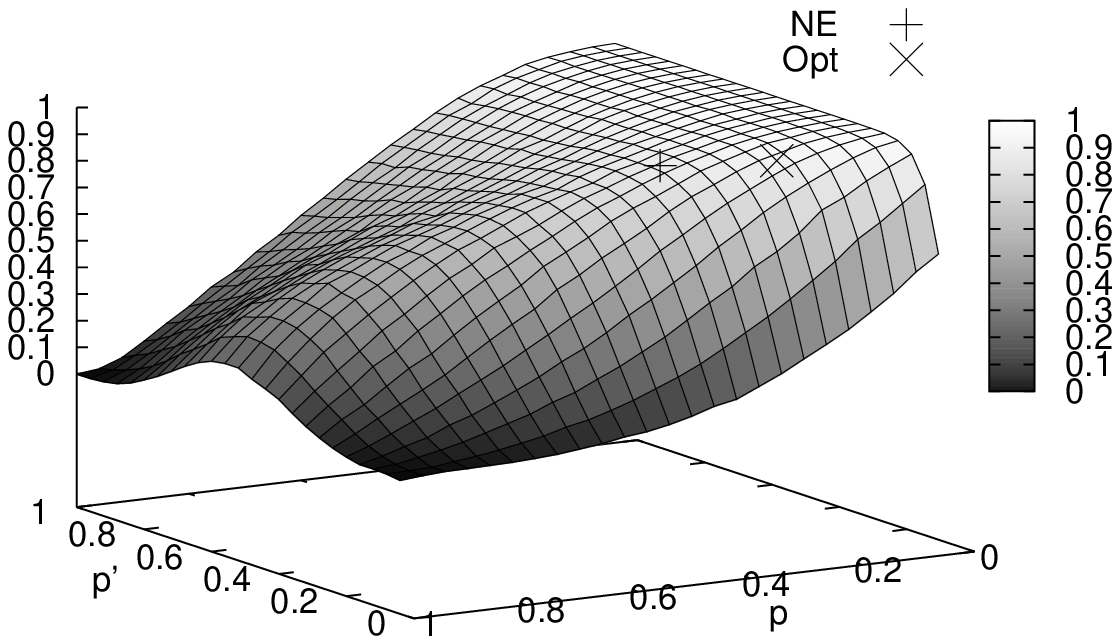} 
  \hspace{\stretch{1}}
  \includegraphics[height=0.25\textheight]{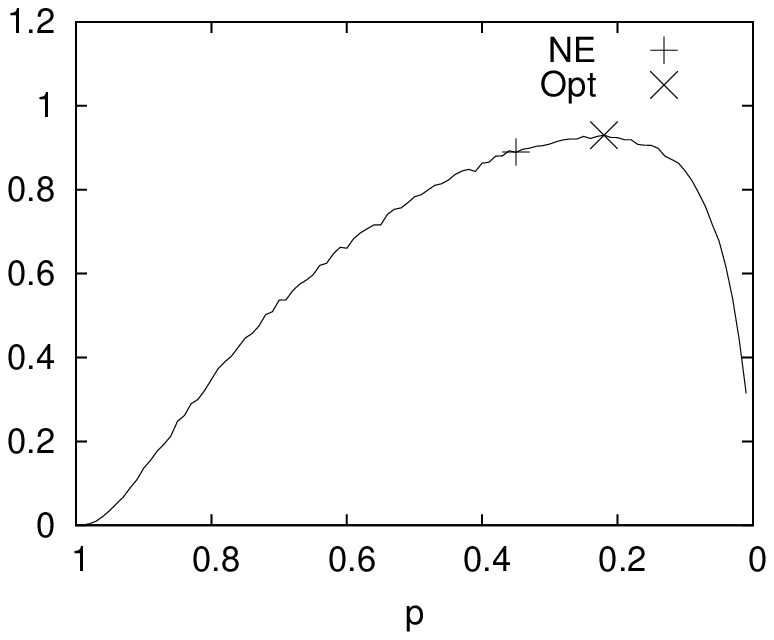}
  \caption{Utility function (left) and its diagonal slice (right) for Aloha with $5$ nodes.}
  \label{fig:alohaplots}
\end{figure}

Fig.~\ref{fig:alohaplots} depicts the plot of the utility function
$U_0(p', p, \dots, p)$ for the network of $5$ nodes for different
values of $p'$ and $p$.  Here $p'$ is a value of \verb|TransmitProb|
of a potentially selfish node, and $p$ is a value for other nodes.
You can also see the computed values of Nash Equlibrium (NE) parameter
and symmetric optimal (Opt) parameter.

Table~\ref{aloha:results} contains the found values of Nash Equilibrium for Aloha with different number of nodes.
The experiments were done on a 8-node cluster, where each node uses Intel(R) Core(TM)2 Quad CPU 2.66GHz processor.
%
%

\subsection{Parameterized Topology for Network Models}

There are situations 
The performance of some network protocols can depend not only on
retransmission parameters as seen previously but also on the actual
topology of the network. In this section we study the impact of different
topologies on the LMAC protocol.

LMAC is a Lightweight Media Access protocol (studied
in~\cite{uppaal_formats,FHM2007}) used for scheduling communication in
wireless sensor networks where the topology is determined by physical
location and radio connectivity of the individual nodes.
One of the goals of the LMAC protocol is to minimize the number of
collisions in the network and to reconfigure the network to avoid further collisions.
The difficulty of studying such protocols stems from the fact that the topology is not known in advance and there are exponentially many topologies (at least $n\cdot 2^{n-1}$ for $n$ nodes with one of them being a gateway), which makes systematic analysis of large networks impractical.
In order to study the robustness of the LMAC protocol against collisions, we propose to examine hundreds of random topologies and then pick and focus on the most problematic ones.
Listing~\ref{lst:lmac-decl} shows how a topology is declared in the
{\sc Uppaal} model: a two-dimensional array of boolean constants gives
the adjacency matrix of the network graph.
The receivers then use the guard {\tt can\_hear[receiver][sender]} when listening for the broadcast channel synchronizations.
\begin{lstlisting}[language={[GUI]Uppaal},caption={Network topology declaration in {\sc Uppaal} model of LMAC.},label={lst:lmac-decl},float={!htb}]
const int NODES = 10;                                     // number of nodes
typedef int[0,NODES@$-$@1] nodeid_t;                         // used to identify node
typedef bool topology_t[nodeid_t][nodeid_t];              // type for topology
const topology_t can_hear = #binarymatrix(NODES, NODES); // adjacency matrix
\end{lstlisting}

In this case we try networks of up to ten nodes and twice as many slots, whereas one slot per node is enough to schedule flawless communication if only nodes were perfectly aware of each others choices.
We used a property $Pr[\Diamond_{time \leq 2000}(col\_count>  42)]$ estimating the probability of having more than 42 collisions after 2000 time units, which hints that there are perpetually reoccurring collisions.

\begin{figure}[!htb]
  \begin{tabular}{@{}c@{}c@{}c@{}c@{}c@{}}
    \includegraphics[height=0.12\textheight]{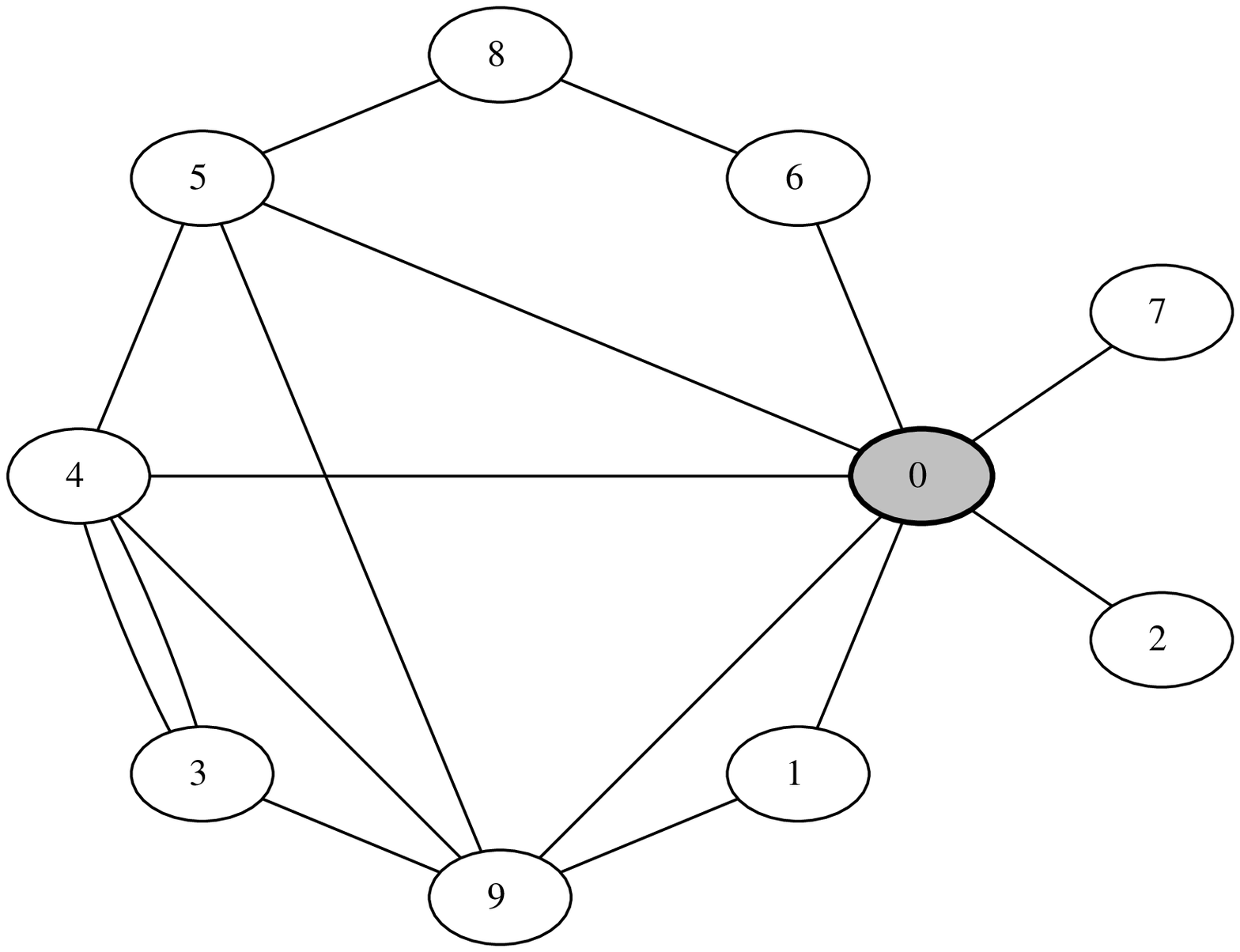} &
    \includegraphics[height=0.12\textheight]{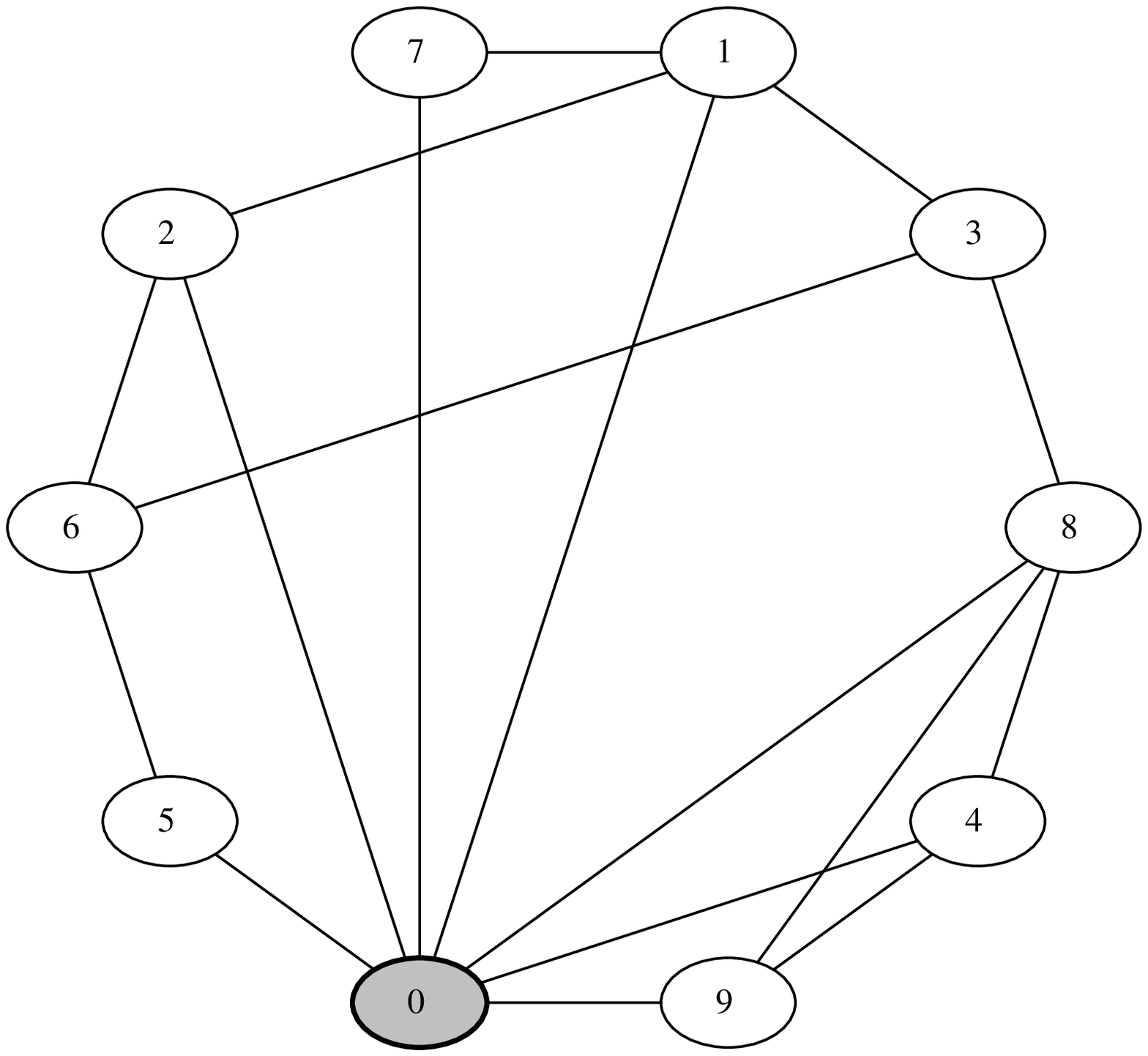} &
    \includegraphics[height=0.12\textheight]{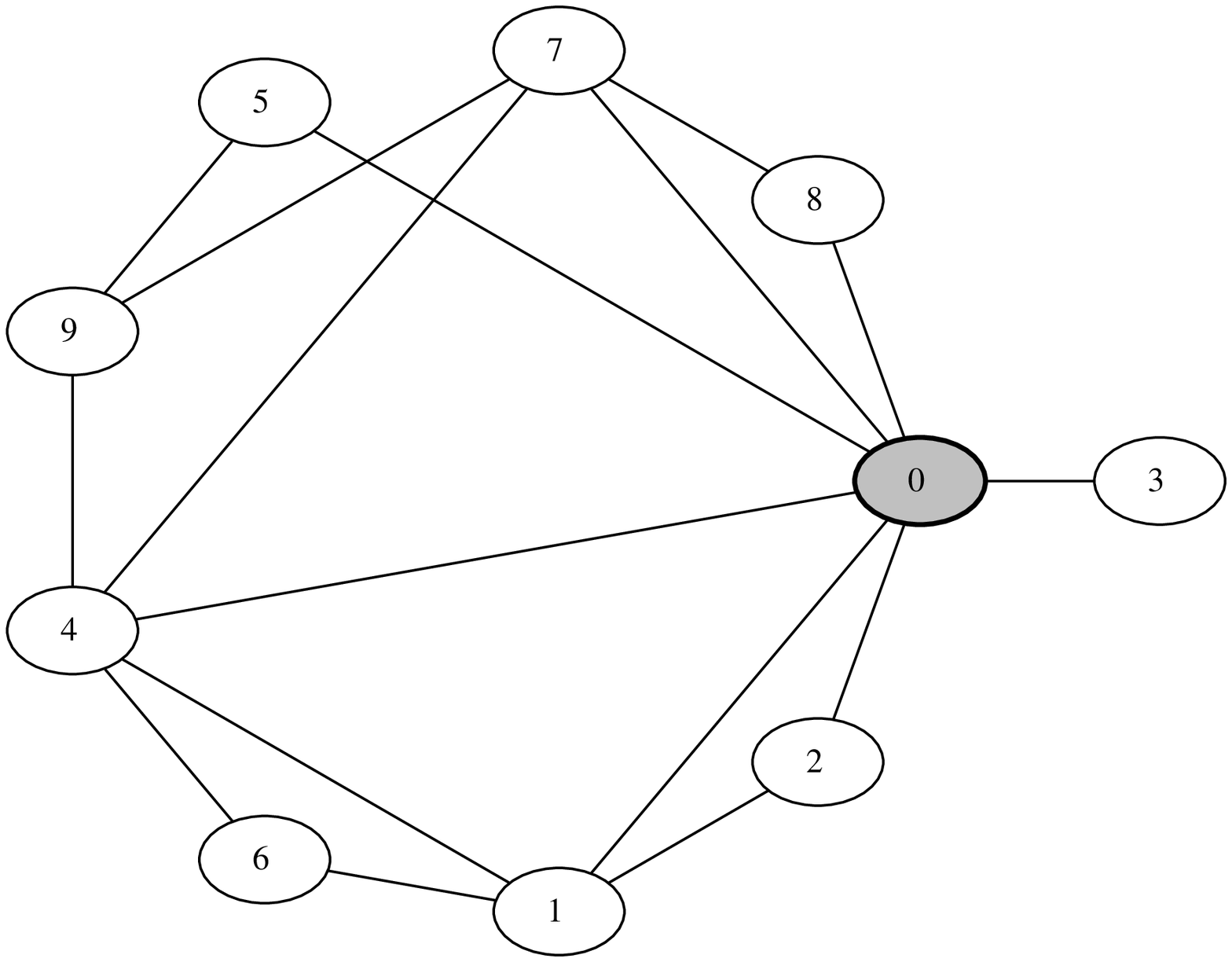} &
    \includegraphics[height=0.12\textheight]{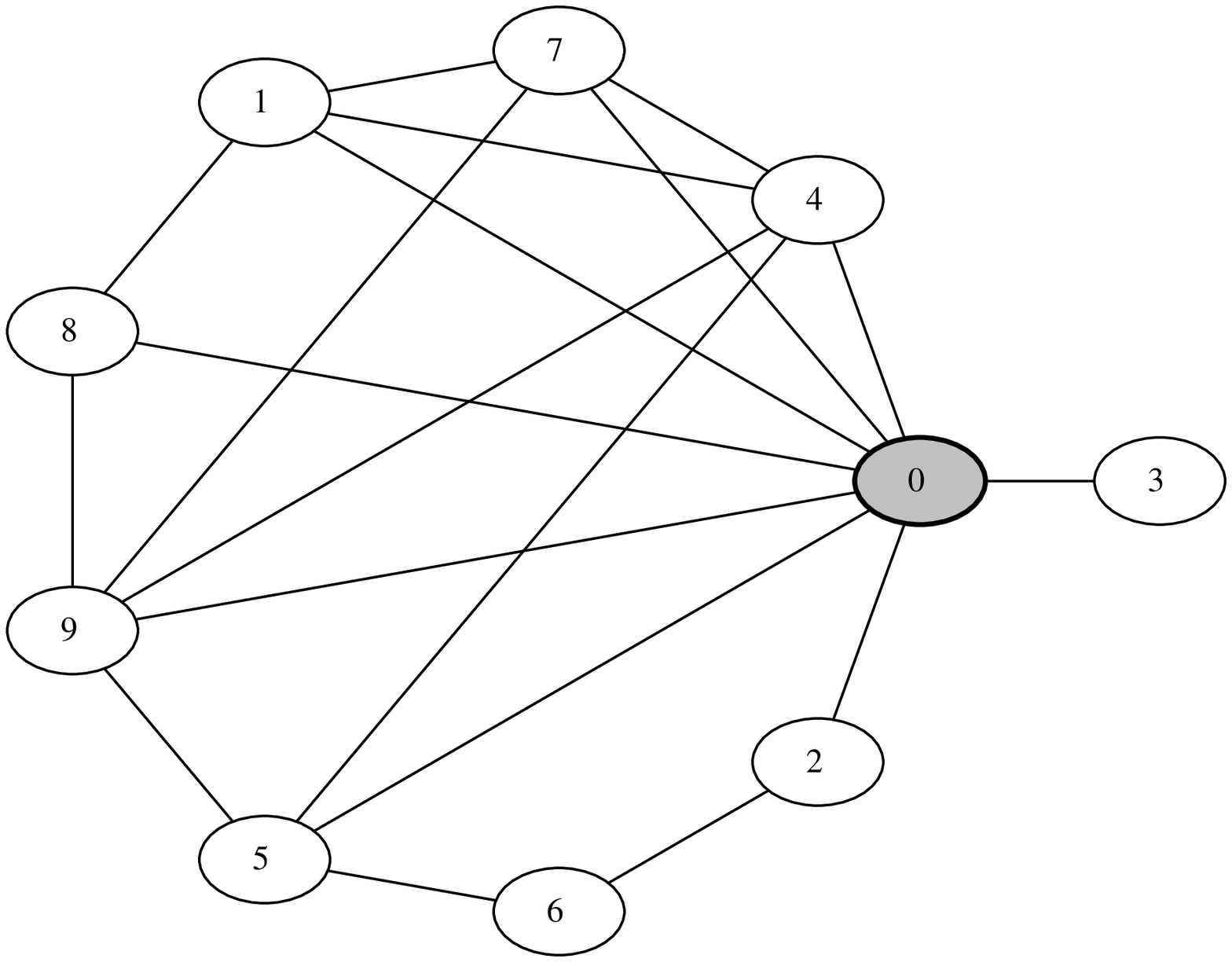} &
    \includegraphics[height=0.12\textheight]{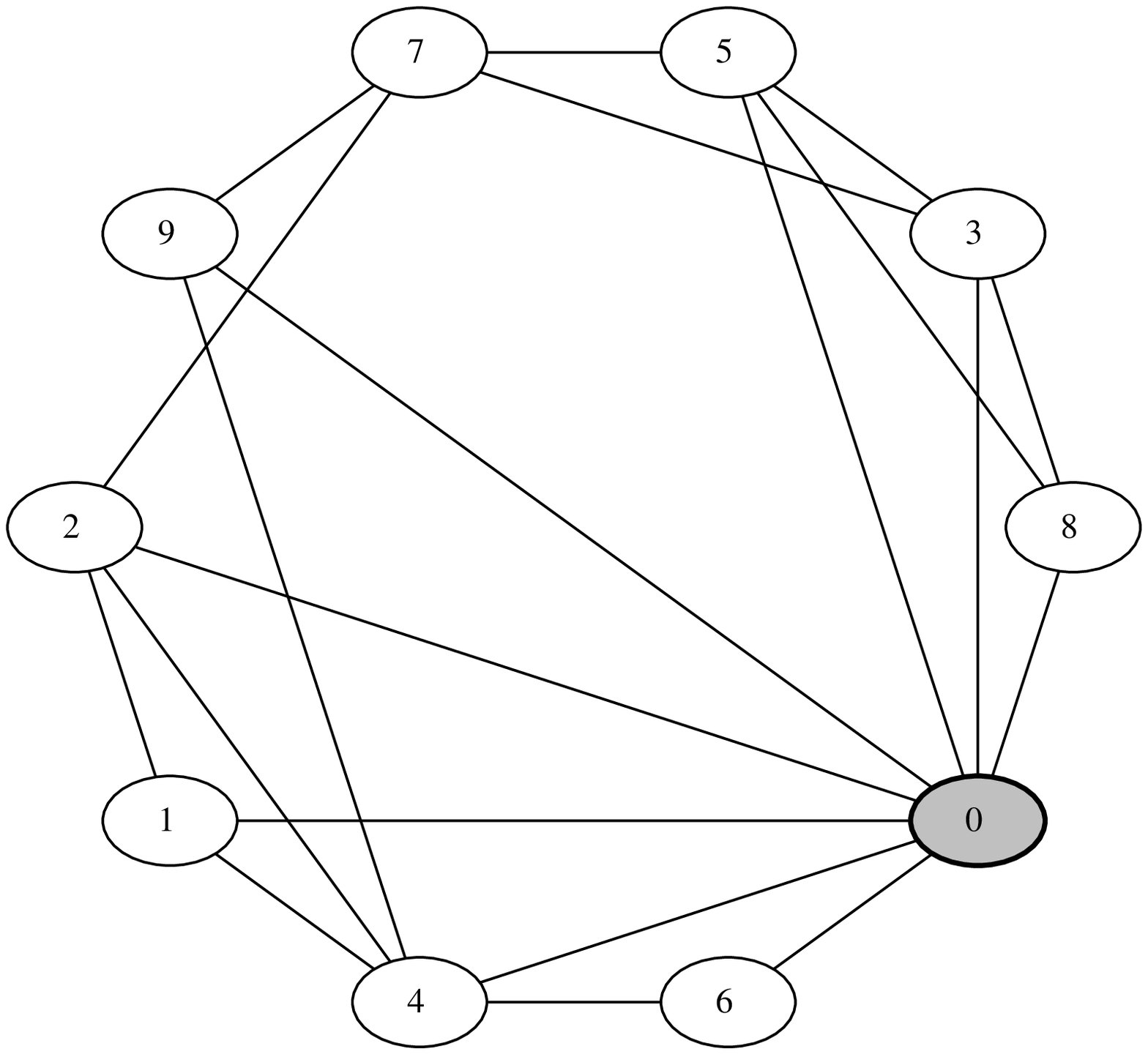} \\
    $p_1=0.630$ & $p_2=0.629$ & $p_3=0.617$ & $p_4=0.607$ & $p_5=0.599$ \\
  \end{tabular}
  \caption{Highest probabilities found by model checking random topologies of 10 nodes.}
  \label{fig:lmac10random}
\end{figure}

The prepared model is then processed by our parametric model-checker that instantiates the keyword {\tt \#binarymatrix} with a concrete random matrix and distributes the verification on a cluster of computers, one instance of the matrix per core. 
Each verification uses \uppaalsmc.
Using the naive randomization, a cluster of 32 cores (the same as in Section~\ref{sec:nash}) can verify 10000 topologies\footnote{We detected 707 duplicates by a post-analysis of the generated instance.} in 6h 50min.
Figure~\ref{fig:lmac10random} shows the five topologies that yield the highest probabilities.
We used low confidence (95\%) statistical parameters to gain performance, thus the estimated probabilities have large $\pm 0.05$ statistical error, but the found topologies can be studied further in \uppaalsmc.

Alternatively we tried generating all graphs up to 10 nodes which are unlikely to be isomorphic.
The procedure is not guaranteed to cover all non-isomorphic classes (it may miss some), but it is very simple and can be recursively described as follows: 
\begin{enumerate}
\item Start with a topology consisting of just one node.
\item Add a new node and consider two new topologies:
  \begin{enumerate}
  \item Connect the new node to all the old nodes, go to step 2 until enough nodes are added.
  \item Leave the new node unconnected at all, go to step 2 until enough nodes are added.
  \end{enumerate}
\item For every node in a topology, make a new topology by marking the node as a gateway.
\item Get rid of the topologies where the gateway is not connected.
\end{enumerate}
Up to the step 2 the procedure generates $2^{n-1}$ topologies which are non-isomorphic for sure, then steps 4 and 5 contain basic heuristics how to pick a gateway, which may yield some isomorphic graphs due to symmetric gateways, but the overhead is small.

\begin{figure}[!htb]
  \centering
  \small
  \begin{tabular}{ccccc}
    \includegraphics[height=0.12\textheight]{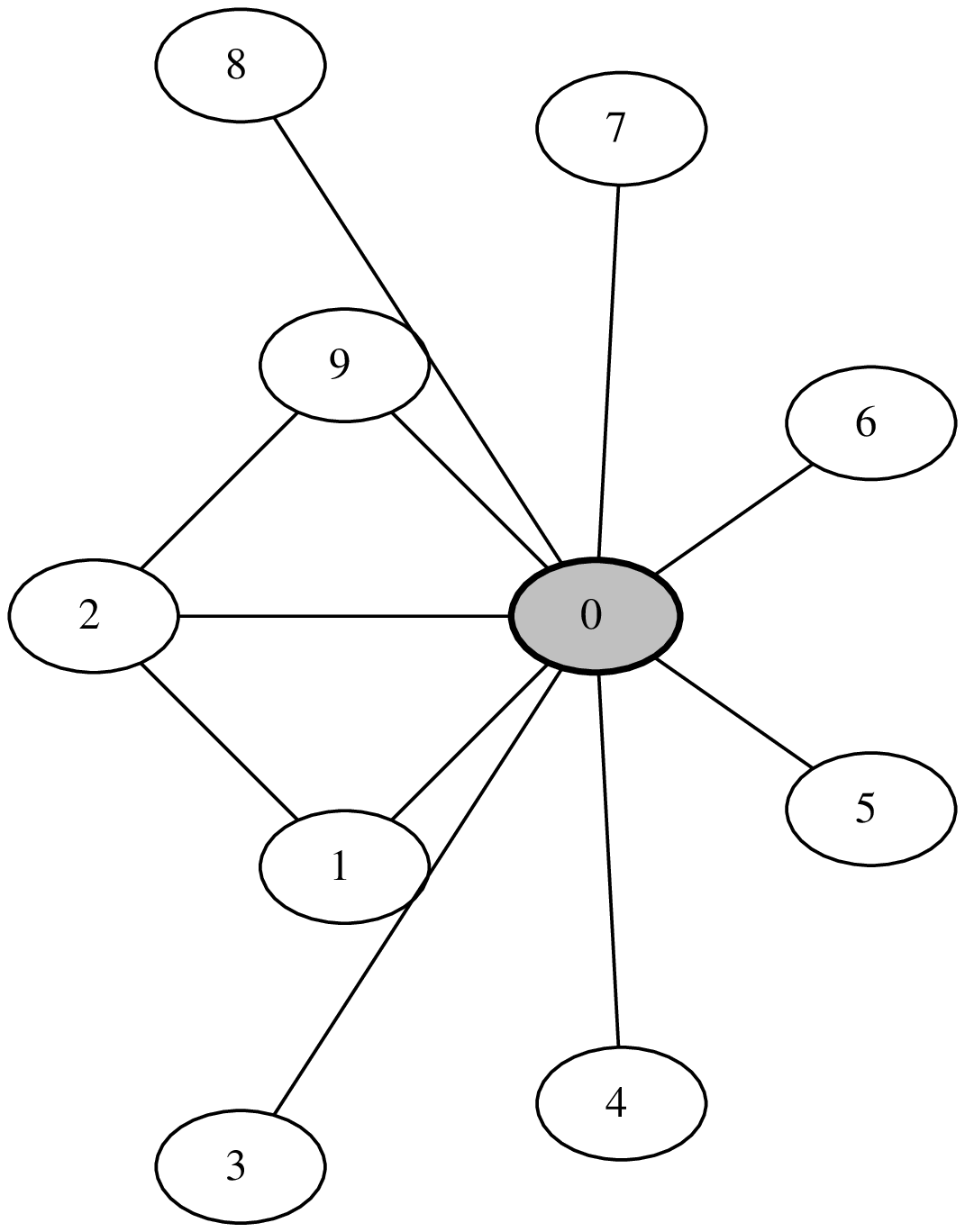} &
    \includegraphics[height=0.12\textheight]{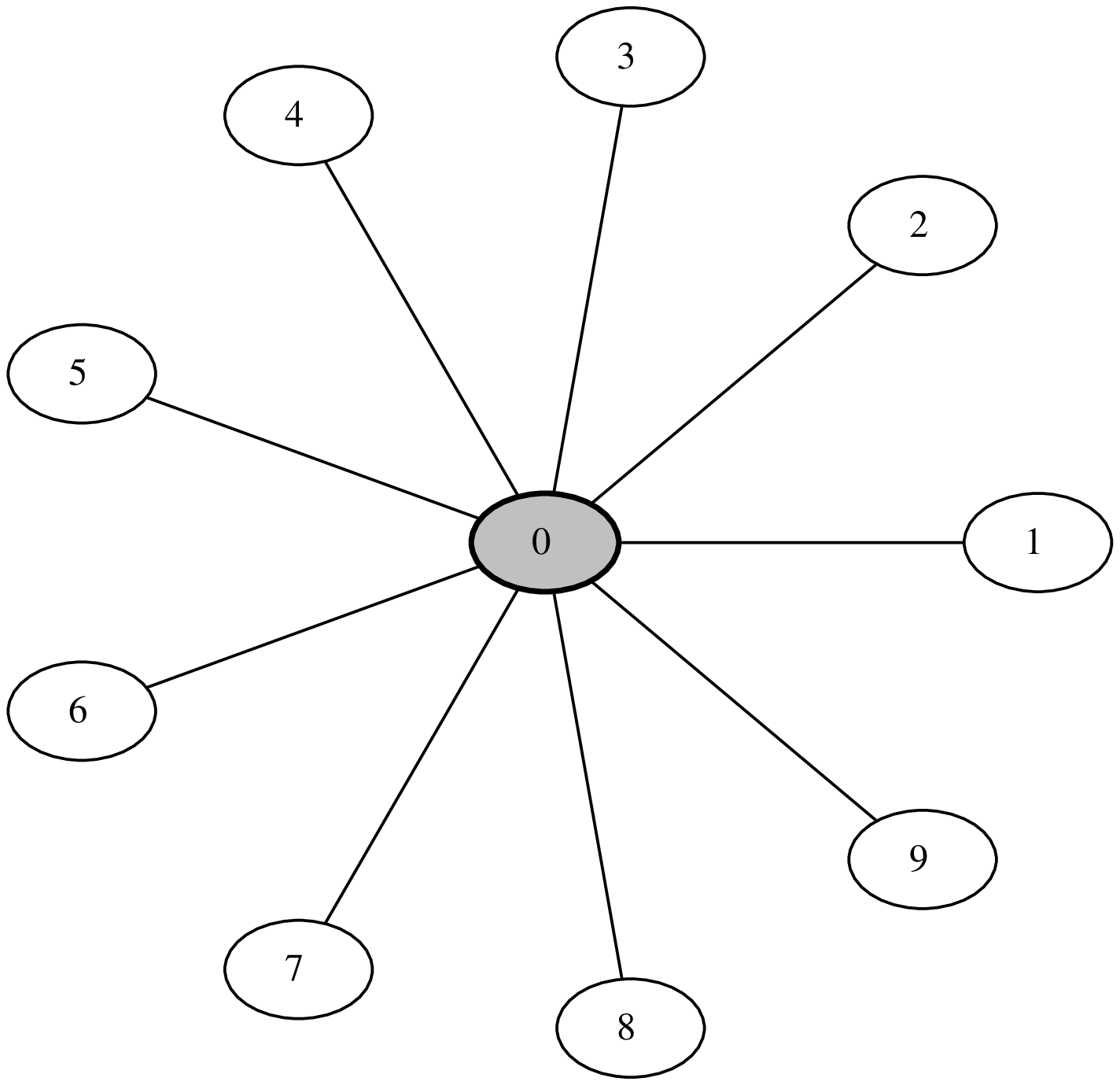} &
    \includegraphics[height=0.12\textheight]{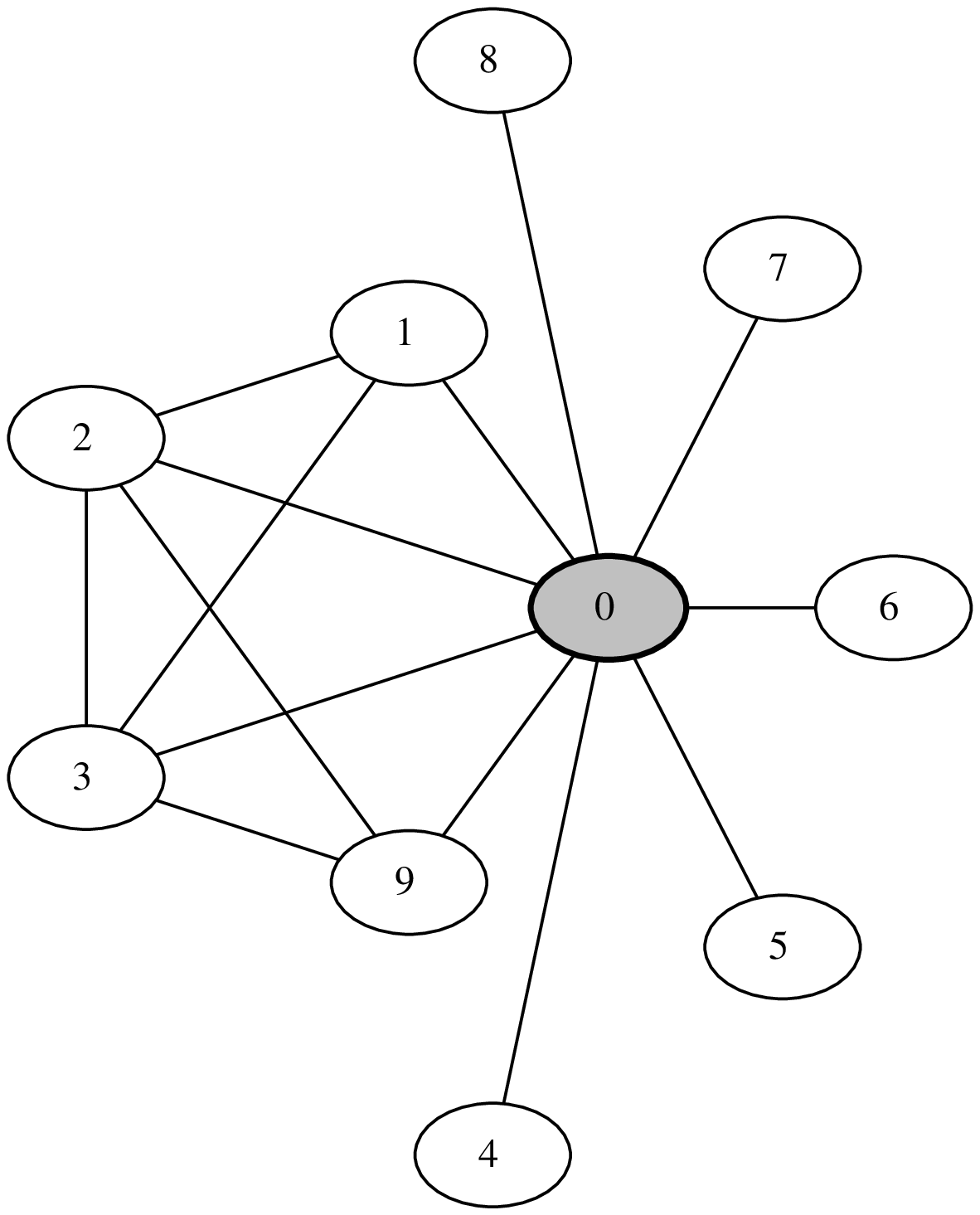} &
    \includegraphics[height=0.12\textheight]{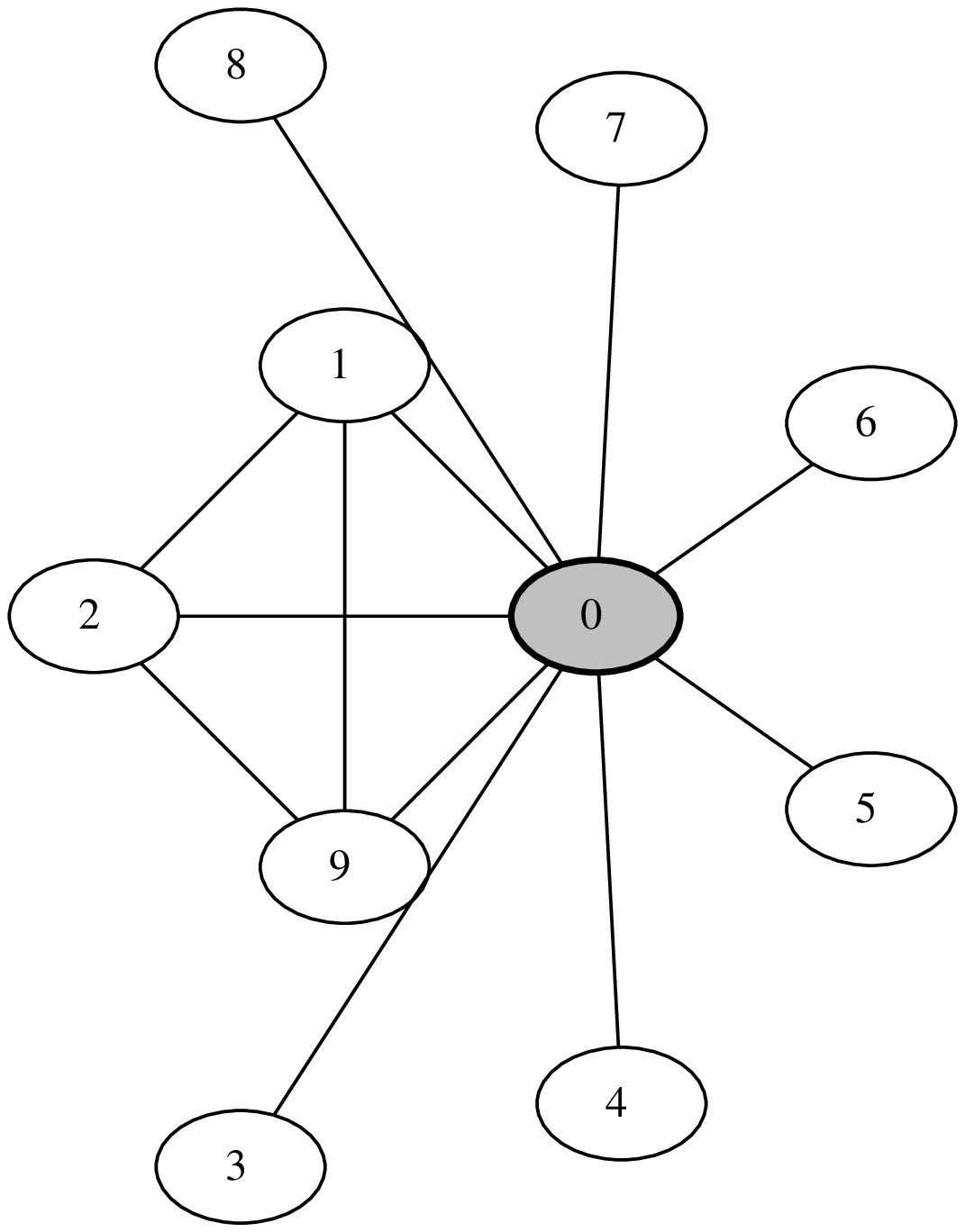} &
    \includegraphics[height=0.12\textheight]{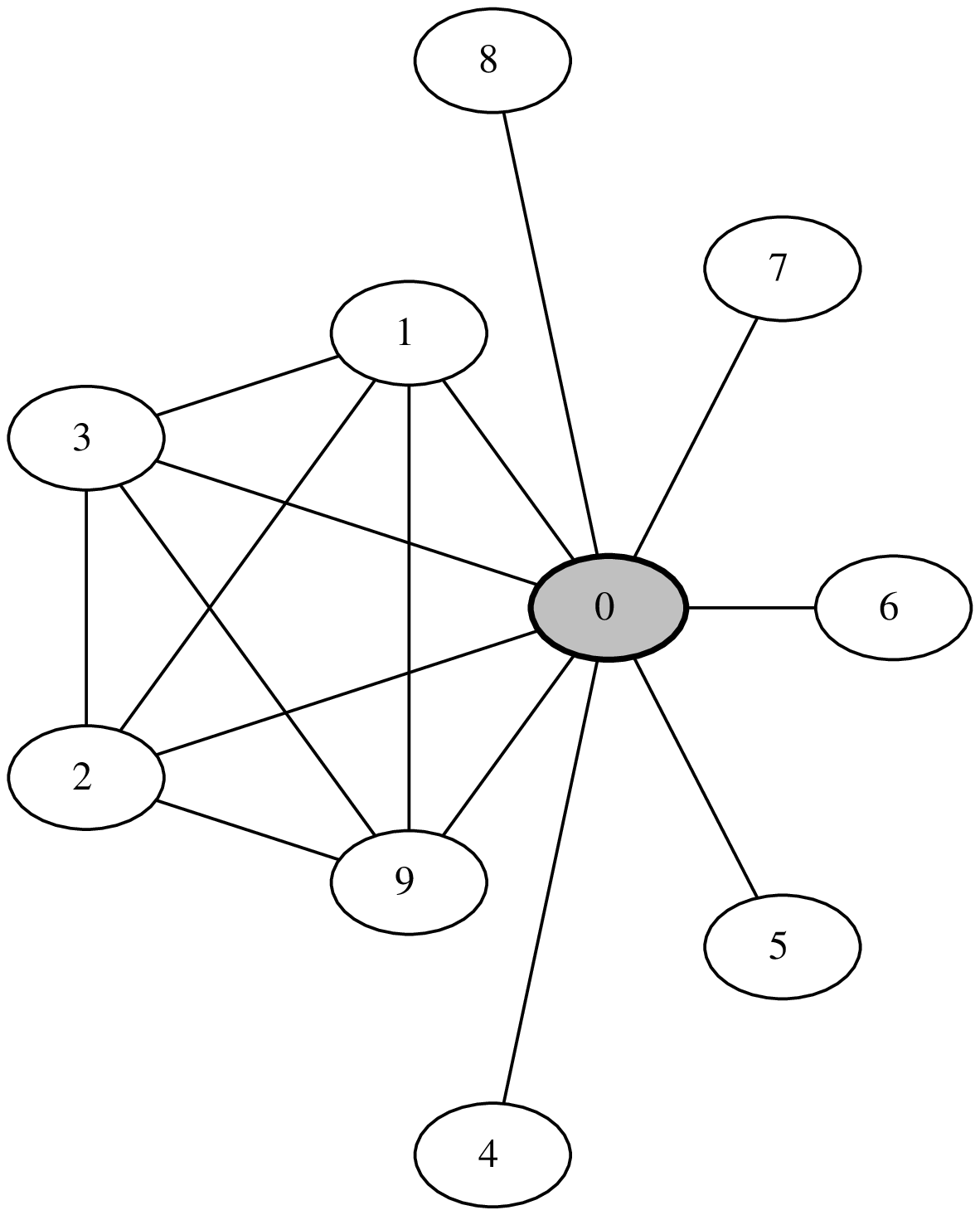} \\
    $p_1=0.919$ & $p_2=0.905$ & $p_3=0.898$ & $p_4=0.894$ & $p_5=0.888$ \\
  \end{tabular}
  \caption{Highest probabilities found by model checking generated topologies of 10 nodes.}
  \label{fig:lmac10smart}
\end{figure}

Figure~\ref{fig:lmac10smart} shows the 5 cases that achieve the
highest probability found by generating 5120 topologies of up to 10 nodes using our heuristics.
The verification took about 3h 30min.
The heuristic procedure has clear advantages over the randomized one but it is not exhaustive. 
On the other hand, the randomized method has the potential to find any topology but without any guarantee.

\section{Conclusion}

This paper proposes new algorithms to distribute statistical model
checking algorithms through a master/slaves architecture. Our results
have been implemented in the UPPAAL SMC toolset. A series of
experiments show that our approach scales better than existing
solutions\,\cite{You05c}. 

As a future work, we will extend our distributed algorithms to the
setting of rare events and unbounded temporal properties. We shall
also implement and distribute Bayesian extensions of the approach we
proposed in \cite{JCLLPZ09}.

\bibliographystyle{eptcs}
\bibliography{references,thesis,s4-biblio,paper}
\end{document}